\documentclass[preprintnumbers, twocolumn, preprintnumbers, PRD, superscriptaddress, nofootinbib]{revtex4-2}

\usepackage{eurosym}
\usepackage{amsfonts}
\usepackage{amsmath}
\usepackage{amssymb}
\usepackage{latexsym}
\usepackage{graphicx}
\usepackage{subfigure}
\usepackage{epstopdf}
\usepackage{xcolor}
\usepackage{soul}
\usepackage{bigints}
\usepackage[colorlinks=true, citecolor=blue, linkcolor=blue, urlcolor=black]{hyperref} 

\renewcommand{\&}{\textup{\symbol{`\&}}}

\begin{document}

\title{{Graviton mass due to dark energy as a superconducting medium -\\
theoretical and phenomenological aspects}}

\author{Nader Inan}
\email{ninan@ucmerced.edu (corresponding author)}

\affiliation{ Clovis Community College, 10309 N. Willow, Fresno, CA 93730 USA}
\affiliation{ Department of Physics, California State University Fresno, Fresno, CA 93740-8031, USA}
\affiliation{ University of California, Merced, School of Natural Sciences, P.O. Box 2039,
Merced, CA 95344, USA}

\author{Ahmed~Farag~{\bf A}li}
\email{aali29@essex.edu; ahmed.ali@fsc.bu.edu.eg}

\affiliation{\footnotesize{Essex County College, 303 University Ave, Newark, NJ 07102, United States.}}
\affiliation{\footnotesize{Department of Physics, Faculty of Science, Benha University, Benha, 13518, Egypt.}}

\author{Kimet {\bf J}usufi}
\email{kimet.jusufi@unite.edu.mk}

\affiliation{\footnotesize{Physics Department, University of Tetova,
Ilinden Street nn, 1200, Tetova, North Macedonia}}

\author{ Abdelrahman Yasser }
\email{ayasser@sci.cu.edu.eg}

\affiliation{\footnotesize{Department of Physics, Faculty of Science, Cairo University, Giza 12613, Egypt}}

\begin{abstract}
\noindent
It is well known that the cosmological constant term in the Einstein field
equations can be interpreted as a stress tensor for dark energy. This stress
tensor is formally analogous to an elastic constitutive equation in
continuum mechanics. As a result, the cosmological constant leads to a
\textquotedblleft shear modulus\textquotedblright\ and \textquotedblleft
bulk modulus\textquotedblright\ affecting all gravitational fields in the
universe. The form of the constitutive equation is also analogous to the
London constitutive equation for a superconductor. Treating dark energy as a
type of superconducting medium for gravitational waves leads to a Yukawa-like gravitational potential and a massive graviton within standard General Relativity.  We discuss a number of resulting phenomenological aspects such as a screening length scale that can also be used to describe the effects generally attributed to dark matter. In addition, we find a gravitational wave plasma frequency, index of refraction, and impedance. The expansion of the universe is interpreted
as a Meissner-like effect as dark energy causes an outward ``expulsion'' of space-time similar to a superconductor
expelling a magnetic field. The fundamental cause of these effects is
interpreted as a type of spontaneous symmetry breaking of a scalar field.
There is an associated chemical potential, critical temperature, and an
Unruh-Hawking effect associated with the formulation.
\end{abstract}

\maketitle

\section{Introduction}
It is well known that the Higgs mechanism in particle physics explains how particles acquire mass. In simple terms, according to the Higgs mechanism, we have to assume the existence of a scalar field known as the Higgs field, which permeates all of the space \cite{Ryder}. This field interacts with particles, and as a consequence of this interaction, particles gain mass. The Higgs mechanism is closely linked to the spontaneous symmetry-breaking process. Specifically, it is now widely believed that symmetry breaking leads to the manifestation of mass in particles where the Higgs field changes from a high-energy state to a low-energy state. In the high-energy state,  particles are massless and the underlying symmetry is unbroken,  while in the low-energy state, particles acquire mass and the symmetry is spontaneously broken.

On the other hand, General Relativity (GR) is the best theory that explains gravitation. Over recent years, this theory has undergone extensive testing, emerging as remarkably successful in its predictions, including the detection of gravitational waves and black hole shadows \cite{EHT1,EHT2,EHT3,EHT4,EHT5,EHT6,EHT7,LIGO}. Assuming the quantum nature of spacetime, the existence of gravitational waves also leads to the prediction of gravitons. Since gravitational waves propagate at the speed of light according to standard GR, then gravitons are expected to be massless.

However, despite the success of GR, there are two open problems in modern cosmology related to the existence of dark matter and dark energy \cite{Salucci:2020nlp}. Specifically, observations hint to the presence of matter that doesn't interact with light but only through the gravitational force. For example, this force is needed to explain the rotating curves in galaxies. However, despite extensive efforts, the elusive dark matter particles that are expected to exist have never been observed in experiments. Meanwhile, in large-scale structures, dark energy plays a pivotal role in driving the acceleration of the universe \cite{Schmidt,Bamba:2012cp}. The nature of dark energy is usually associated with the quantum vacuum energy \cite{Carroll:2000fy}. However, this leads to the \textquotedblleft dark energy problem\textquotedblright\ due to the discrepancy between the predicted and observed energy density of the vacuum. From the theoretical point of view, we can calculate the vacuum energy from the quantum fluctuations, however, the observed value of vacuum energy, inferred from the accelerated expansion of the universe, appears to be drastically smaller than the value predicted by theoretical calculations. This mismatch continues to be an open and unsolved problem in modern cosmology.

Dark energy is effectively encoded in the Einstein field equations by the cosmological constant $\Lambda$. In fact, both dark energy and dark matter are currently incorporated into the Einstein field equations simply by hand, as their existence are inferred to exist primarily based on observational data since no definitive underlying mechanism for their nature has been identified to date. Some ideas to explain the late-time cosmic acceleration have been explored. In particular, some ideas use a scalar field with non-zero potentials called a quintessence field \cite{Peebles,Caldwell:1997ii}. An important aspect to be noted is that contrary to the cosmological constant, the quintessence field can change with time \cite{Olivares:2007rt}. 

Apart from the idea that dark matter is described by a particle, the possibility that dark matter might emerge as a modified law of gravity was studied extensively. For example, Modified Newtonian Dynamics (MOND) was proposed by Milgrom \cite{Milgrom1}. In particular, Newton's force law is modified on large distance scales leading to an apparent effect that mimics dark matter. 

However, it might be argued that the simplest and most natural possibility is to modify GR by adding mass to the graviton. Massive gravity has a long history, dating back to the work of Fierz and Pauli \cite{Fierz}. However, it was believed for a long time that the graviton mass must be zero due to the presence of a discontinuity between the massless and massive theories \cite{ Zakharov}. This problem was addressed by Vainstein \cite{Vainstein} who found that there exists a scale below which the massive graviton behaves like a massless particle and therefore the graviton could have a small nonzero mass. Another issue related to massive gravity was found by Deser and Boulware \cite{Boulware} showing that the massive theory is ill behaved because in addition to the five degrees of freedom of the massive graviton, there must be an extra scalar degree of freedom which does not decouple. Hence the massive gravity theory leads to instabilities (the emergence of the Boulware-Deser ghost field).  However, it was shown recently that one can generate mass for the graviton via the Higgs mechanism for gravitons which was proposed in an interesting work \cite{Chamseddine:2010ub} where the authors used four scalars with global Lorentz symmetry and showed that in the broken symmetry phase, the graviton absorbs all scalar fields and we end up with a theory of a massive spin-2 particle with five degrees of freedom and free of ghosts.  Another important step forward was made by de Rham, Gabadadze, and Tolley \cite{deRham:2010kj} who found a theory of massive gravity that is free from instabilities and ghosts. Furthermore, it was found that other modified gravity theories such as bigravity theories can be ghost-free theories \cite{Hassan:2011zd}. Very recently, the phenomenological aspects of Yukawa-like corrections due to massive gravitons were studied in the context of cosmology where dark matter is obtained by means of the long-range interaction via a Yukawa modified potential \cite{Jusufi:2023xoa,Jusufi:2024ifp,Gonzalez:2023rsd,Jusufi:2024ifp}.\\
\indent In this paper, we propose the novel idea of a Higgs-like mechanism for gravitons analogous to a superconductor generating mass for photons. In particular, dark energy is modeled as an effective superconducting medium.  The analogy is inspired from a quantum gravity point of view. Due to the discreteness of spacetime at the Planck length scale, one can model dark energy as a superconductor where the role of the atomic lattice is essentially played by ``spacetime atoms.'' \cite{Spacetimeatoms1, Spacetimeatoms2}. The dynamics of such a structure could lead to a phonon-graviton interaction where ``phonons'' occur in the lattice of ``spacetime atoms'' just as phonons exist in the ionic lattice of a superconductor. The idea of modeling dark energy as a superconductor also appears in \cite{Liang:2015tfa}, however, the authors use a scalar-vector-tensor gravity model rather than using standard GR and arguing that the mass of the graviton emerges from a spontaneous symmetry breaking mechanism. Moreover, we show many interesting phenomenological implications of this model such as a gravitational penetration depth (screening length scale), plasma frequency, index of refraction, and impedance. We also show that the expansion of the universe can be understood in terms of a Meissner-like effect as dark energy causes an outward ``expulsion'' of space-time similar to a superconductor expelling a magnetic field. 

\section{Photon mass and spontaneous symmetry breaking in a superconductor}

We begin with a general review of the method for determining the mass and
length scale of a scalar field, and the associated mechanism of spontaneous
symmetry breaking. Recall that the Klein-Gordon equation (in flat
space-time) is%
\begin{equation}
\square \varphi -k_{\text{c}}^{2}\varphi =0  \label{K-G (flat)}
\end{equation}%
where $k_{\text{c}}\equiv mc/\hslash $ is the reduced Compton wave number, $%
\varphi $ is a scalar field, and $\square \equiv -\tfrac{1}{c^{2}}\partial
_{t}^{2}+\nabla ^{2}$ is the d'Alembert operator in flat space-time. The
second term of $\left( \ref{K-G (flat)}\right) $ is considered the
\textquotedblleft mass term\textquotedblright\ since setting $m=0$ would
lead to $\square \varphi =0$ which is the wave equation for a massless
scalar field in vacuum. Therefore, the mass of the scalar field is
identified as%
\begin{equation}
m=k_{\text{c}}\hslash /c  \label{mass scale}
\end{equation}%
Also note that this mass term has an associated length scale given by the
reduced Compton wavelength, $\lambdabar =1/k_{\text{c}}$ which gives%
\begin{equation}
\lambdabar =\frac{\hslash }{mc}  \label{length scale}
\end{equation}%
Therefore, for a massive scalar field, $\left( \ref{mass scale}\right) $ and 
$\left( \ref{length scale}\right) $ give a general form for determining the
mass and length scale, respectively, of a particular field theory. For
example, we can apply this concept to the case of electromagnetic fields in
a superconductor. The covariant London constitutive equation for a
superconductor is $J^{\mu }=-\Lambda _{\text{L}}A^{\mu }$, where $J^{\mu }$
is the electric four-current, $A^{\mu }$ is the four-potential, $\Lambda _{%
\text{L}}\equiv n_{\text{s}}e^{2}/m_{\text{e}}$ is the London constant, and $%
n_{\text{s}}$ is the number density of Cooper pairs that effectively form a
condensate and undergo dissipationless acceleration in response to an
electromagnetic field. In the Lorenz gauge, $%
\partial _{\mu }A^{\mu }=0$, Maxwell's equation becomes $\square A^{\mu
}=-\mu _{0}J^{\mu }$. Using the covariant London constitutive equation leads
to%
\begin{equation}
\square A^{\mu }-k_{\text{EM}}^{2}A^{\mu }=0  \label{EM wave}
\end{equation}%
where an effective Compton wave number for the electromagnetic field is
defined as%
\begin{equation}
k_{\text{EM}}^{2}\equiv \frac{\mu _{0}n_{\text{s}}e^{2}}{m_{\text{e}}}
\label{k_EM}
\end{equation}%
Comparing $\left( \ref{EM wave}\right) $ to $\left( \ref{K-G (flat)}\right) $%
, it is evident that the second term in $\left( \ref{EM wave}\right) $ is
effectively a \textquotedblleft mass term\textquotedblright\ which implies
that the vector potential can be viewed as a massive vector field in the
superconductor. In \cite{Ryder} this is interpreted as the photon
developing a mass within the superconductor.\bigskip 

Using $\lambda _{\text{EM}}\equiv 1/k_{\text{EM}}$, we find that the length
scale associated with the Compton wave number is%
\begin{equation}
\lambda _{\text{EM}}=\sqrt{\frac{m_{\text{e}}}{\mu _{0}n_{\text{s}}e^{2}}}
\label{lamda_L}
\end{equation}%
This is the well-known London penetration depth which characterizes the
Meissner effect within a superconductor. For Niobium\footnote{%
We consider that each atom contributes two conduction electrons and only $%
10^{-3}$ of the conduction electrons are in a superconducting state, so that 
$n_{\text{s}}\approx 2n\left( 10^{-3}\right) $, where $n=\rho _{\text{m}}/m$
is the number density of atoms. For Niobium, the mass density is $\rho _{%
\text{m}}\approx 8.6\times 10^{3}$ kg/m$^{3}$ and the mass per atom is $%
m\approx 1.5\times 10^{-25}$ kg/atom. Then the number density of atoms is $%
n\approx 5.7\times 10^{28}$ m$^{-3}$ and therefore the number density of
Cooper pairs is $n_{\text{s}}\approx 2n\left( 10^{-3}\right) \approx
1.1\times 10^{26}~$m$^{-3}$.}, this comes out to a theoretical value of
approximately 500 nm for the penetration depth. (The actual measured value
of the penetration depth for Niobium is $\lambda _{\text{EM}}\approx 39$
nm.) Furthermore, using $\left( \ref{mass scale}\right) $ leads to a photon
mass given by%
\begin{equation}
m_{\text{photon}}=\frac{k_{\text{EM}}\hslash }{c}=\dfrac{\hslash }{c}\sqrt{%
\dfrac{\mu _{0}n_{\text{s}}e^{2}}{m_{\text{e}}}}
\end{equation}%
We can also write this in terms of the London penetration depth as $m_{\text{%
photon}}=\dfrac{\hslash }{c\lambda _{\text{EM}}}$. Using the known value of $%
\lambda _{\text{EM}}\approx 39$ nm for Niobium leads to $m_{\text{photon}%
}\approx 9.0\times 10^{-36}$ kg for the photon mass in a Niobium
superconductor.\footnote{%
See \cite{Tu} for various limits on measured photon mass as well as other
physical implications of a massive photon.}\bigskip 

The photon becoming massive within a superconductor can also be understood
as a spontaneous symmetry breaking of the electromagnetic gauge symmetry.%
\footnote{%
Superconductivity can be thought of as a condensed-matter analog
of the Higgs phenomena, in which a condensate of Cooper pairs of electrons
spontaneously breaks the U(1) gauge symmetry of electromagnetism. For more
details, see Sections 8.3-8.4 of Ryder's text \cite{Ryder}, 21.6 of
Weinberg's text \cite{WeinbergQuantumTheory(V2)}, and 11.7-11.9 of
Griffiths' text \cite{GriffithsIntrotoParticles}. For an insightful
alternative perspective on this point, see Greiter \cite{Greiter}.} Recall
that a gauge transformation in electromagnetism is given by $A^{\mu
}\rightarrow A^{\mu }+\partial ^{\mu }\chi $, where $\chi $ is the
associated gauge generating function. However, in a superconductor, there is
essentially a preferred gauge, namely, the London gauge given by $\nabla
\cdot \vec{A}=0$. This gauge choice is implicit in the use of the London
constituent equation, $\vec{J}=-\Lambda _{\text{L}}\vec{A}$. To observe
this, we recognize that $n_{\text{s}}$ is a constant in time throughout the
superconductor and therefore the charge density of Cooper pairs, $\rho =2n_{%
\text{s}}e$, must also be a constant.\footnote{%
The condition $\dot{\rho}=0$ is necessitated by the demand of local charge
neutrality, i.e., that the positive ionic charge within each unit cell of
the ionic lattice must be exactly balanced by the negative charge density of
the Cooper pairs within the same unit cell. Otherwise, the Coulomb energy
due to any slight charge imbalance would be huge and would drive the
superconductor back to a state of electrostatic equilibrium in which charge
neutrality is immediately re-established.} Since $\dot{\rho}=0$, then by the
continuity equation, $\dot{\rho}=\nabla \cdot \vec{J}=0$. By the
London constitutive equation, $\vec{J}=-\Lambda _{\text{L}}\vec{A}$, it
follows that $\nabla \cdot \vec{A}=0$.\bigskip 

Lastly, we point out that the concept of a massive photon can also be
observed by considering a phi-fourth theory Lagrangian density given in
equation (8.36) of \cite{Ryder} as%
\begin{equation}
\mathcal{L}=\left( D^{\mu }\varphi \right) ^{\ast }\left( D_{\mu }\varphi
\right) +\alpha \left\vert \varphi \right\vert ^{2}+\frac{\beta }{2}%
\left\vert \varphi \right\vert ^{4}-\dfrac{1}{4\mu _{0}}F^{\mu \nu }F_{\mu
\nu }  \label{phi-fourth theory Lagrangian density}
\end{equation}%
where $D_{\mu }=\partial _{\mu }-\tfrac{ie}{\hslash }A_{\mu }$, and $\varphi 
$ is a complex scalar field. According to \cite{Ryder}, this is related to
the Ginzburg-Landau free energy density, where $\alpha \sim \left( T-T_{%
\text{c}}\right) $ near the critical temperature $T_{\text{c}}$, and $%
\varphi $ is the macroscopic many-particle wave function, with its use
justified by the Bardeen-Cooper-Schrieffer (BCS) theory. Notice that
multiplying out terms in the Lagrangian density will lead to a term
involving $A^{\mu }A_{\mu }\left\vert \phi \right\vert ^{2}$. This term is
understood to show that the photon has become massive since it corresponds
to $k_{\text{EM}}^{2}A^{\mu }$ in $\left( \ref{EM wave}\right) $, where $k_{%
\text{EM}}=m_{\text{photon}}c/\hslash $.\bigskip 

Similarly, \cite{Hinterbichler} shows that the Fierz-Pauli action \cite%
{Fierz} describes a massive spin-2 particle in flat space-time,
carried by a symmetric tensor field $h_{\mu \nu }$. The action appears as%
\begin{eqnarray}
S &=&\int d^{D}x\left[ -\frac{1}{2}\partial _{\lambda }h_{\mu \nu }\partial
^{\lambda }h^{\mu \nu }+\partial _{\mu }h_{\nu \lambda }\partial ^{\nu
}h^{\mu \lambda }-\partial _{\mu }h^{\mu \nu }\partial _{\nu }h\right.  
\notag \\
&&\qquad \left. +\frac{1}{2}\partial _{\lambda }h\partial ^{\lambda }h-\frac{%
1}{2}m^{2}\left( h_{\mu \nu }h^{\mu \nu }-h^{2}\right) \right] 
\label{FP Lagrangian}
\end{eqnarray}%
Note that when $m=0$, the remaining terms in $\left( \ref{FP Lagrangian}%
\right) $ match those obtained by linearizing the Einstein-Hilbert action
which describes a spin-2 \textit{massless} graviton. (This is analogous to
the standard Maxwell Lagrangian density describing a massless photon.)
However, the last term in $\left( \ref{FP Lagrangian}\right) $ is a graviton
mass term analogous to the photon mass term described by $A^{\mu }A_{\mu
}\left\vert \phi \right\vert ^{2}$ in $\left( \ref{phi-fourth theory
Lagrangian density}\right) $. This implies that massive gravity requires a
modification to the standard Einstein-Hilbert action. However, we show in
this paper that a massive graviton can be obtained without modifying the
Einstein-Hilbert action. Rather, the graviton mass emerges naturally from
the Einstein field equations due to the presence of the Cosmological
Constant.\vspace*{0.3in}

\section{Constitutive equations from continuum mechanics}

Einstein's field equation of GR can be written as%
\begin{equation}
G_{\mu \nu }+\Lambda g_{\mu \nu }=\kappa T_{\mu \nu }  \label{EFE}
\end{equation}%
where $\kappa \equiv 8\pi G/c^{4}$, and $\Lambda $ is the cosmological
constant. The term with $\Lambda g_{\mu \nu }$ can be taken to the right
side and used to define a dark energy stress tensor. This leads to
\begin{equation}
G_{\mu \nu }=\kappa \left( T_{\mu \nu }^{\text{matter}}+T_{\mu \nu }^{\text{%
DE}}\right) 
\end{equation}%
where the dark energy stress tensor is defined as $T_{\mu \nu }^{\text{DE}%
}\equiv -\frac{\Lambda }{\kappa }g_{\mu \nu }$. The metric can be written in
terms of a perturbation to flat space-time: $g_{\mu \nu }=\eta _{\mu \nu
}+h_{\mu \nu }$. Then the dark energy stress tensor becomes
\begin{equation}
T_{\mu \nu }^{\text{DE}}=-\frac{\Lambda }{\kappa }\eta _{\mu \nu }-\frac{%
\Lambda }{\kappa }h_{\mu \nu }  \label{T_mu,nu^DE}
\end{equation}%
The first term on the right side of $\left( \ref{T_mu,nu^DE}\right) $ is a
stress tensor that is inherent to the existence of dark energy even when $%
h_{\mu \nu }=0$. This is analogous to the inherent energy density of a
superconductor which the London brothers described as \textquotedblleft
frozen in\textquotedblright\ in their famous work on superconductors \cite%
{London}. In other words, this energy density is present even when the system is unperturbed by a field.\bigskip 

However, the second term of $\left( \ref{T_mu,nu^DE}\right) $ is essentially
a constitutive equation that describes the stress produced by the presence
of a gravitational field, $h_{\mu \nu }$, that perturbs the system. It is
effectively an interaction term between a gravitational field and dark
energy, where dark energy is viewed as a \textquotedblleft
medium.\textquotedblright\ In fact, this interaction term can be written as%
\begin{equation}
T_{\mu \nu }^{\left( \text{DE}\right) \text{~interaction}}=-\frac{\Lambda }{%
\kappa }h_{\mu \nu }  \label{T_mu,nu^DE (interaction)}
\end{equation}%
Notice how $\left( \ref{T_mu,nu^DE (interaction)}\right) $ has the same form
as the covariant London constitutive equation for a superconductor: $J^{\mu
}=-\Lambda _{\text{L}}A^{\mu }$. In fact, $\left( \ref{T_mu,nu^DE
(interaction)}\right) $ could be referred to as a \textquotedblleft
gravito-London constitutive equation.\textquotedblright \bigskip 

The physical meaning of $\left( \ref{T_mu,nu^DE (interaction)}\right) $ can
also be understood in the context of continuum mechanics which provides a
constitutive equation relating stress and material strain, $U_{ij}$. In
equation (4.6)\ of \cite{LandauandLifshitz(elastic)}, a constitutive equation is written as%
\begin{equation}
T_{ij}=\tfrac{1}{3}BU\delta _{ij}+\mu \left( U_{ij}-\tfrac{1}{3}U\delta
_{ij}\right)   \label{T_ij (material)}
\end{equation}%
where $B$ is the bulk modulus, $\mu$ is the shear modulus, and $U\equiv \delta
^{ij}U_{ij}$. Taking the trace of $\left( \ref{T_ij
(material)}\right) $ gives%
\begin{equation}
T=BU  \label{trace}
\end{equation}%
where $T\equiv \delta ^{ij}T_{ij}$. Similarly, using $\left( \ref{T_mu,nu^DE}%
\right) $ to evaluate $T^{\text{DE}}=\delta ^{ij}T_{ij}^{\text{DE}}$ gives%
\begin{equation}
T^{\text{DE}}=-\frac{3\Lambda }{\kappa }-\frac{\Lambda }{\kappa }H
\label{trace of T_mu,nu^DE}
\end{equation}%
where $H\equiv \delta ^{ij}h_{ij}$. Comparing the stress found in the 
\textit{material} constitutive equation $\left( \ref{trace}\right) $, and
the \textit{gravitational} constitutive equation $\left( \ref{trace of
T_mu,nu^DE}\right) $, shows that there is effectively a \textquotedblleft
gravitational bulk modulus\textquotedblright\ given by\footnote{%
The minus sign difference between $\left( \ref{trace}\right) $ and $\left( %
\ref{trace of T_mu,nu^DE}\right) $ is due to the fact that $\left( \ref%
{trace}\right) $ describes an \textit{external} stress $\left( T_{\text{%
external}}\right) $ causing an \textit{internal} material strain $\left(
U\right) $, whereas $\left( \ref{trace of T_mu,nu^DE}\right) $ describes an 
\textit{external} gravitational strain $\left( H\right) $ causing an \textit{%
internal} stress $\left( T_{\text{internal}}\right) $. The cause and effect
of these equations is reversed. In fact, mechanical equilibrium requires the 
\textit{internal} stress to be equal and opposite to the \textit{external}
stress. Hence $T_{\text{external}}=-T_{\text{internal}}$ which explains the
minus sign difference between $\left( \ref{trace}\right) $ and $\left( \ref%
{trace of T_mu,nu^DE}\right) $, as well as between $\left( \ref{T_ij
(material)^TT}\right) $ and $\left( \ref{T_ij (DE)^TT}\right) $.}%
\begin{equation}
B_{\text{G}}\equiv \frac{\Lambda }{\kappa }
\end{equation}%
This quantity determines the bulk \textquotedblleft
stiffness\textquotedblright\ of dark energy in response to a longitudinal
gravitational strain field.\bigskip 

Furthermore, the transverse-traceless spatial stress tensor can be written as%
\begin{equation}
T_{ij}^{\text{TT}}\equiv T_{ij}-\tfrac{1}{3}T\delta _{ij}  \label{T_ij^TT}
\end{equation}%
where $\partial _{i}T_{ij}^{\text{TT}}=0$ (to be transverse) and $\delta
^{ij}T_{ij}^{\text{TT}}=0$ (to be traceless). Inserting $\left( \ref{T_ij
(material)}\right) $ and $\left( \ref{trace}\right) $ into $\left( \ref%
{T_ij^TT}\right) $ gives%
\begin{equation}
T_{ij}^{\text{TT}}=\mu U_{ij}^{\text{TT}}  \label{T_ij (material)^TT}
\end{equation}%
where the transverse-traceless material strain is $U_{ij}^{\text{TT}}\equiv
U_{ij}-\tfrac{1}{3}U\delta _{ij}$. Similarly, inserting $\left( \ref%
{T_mu,nu^DE}\right) $ into $\left( \ref{T_ij^TT}\right) $ leads to%
\begin{equation}
T_{ij}^{\left( \text{DE}\right) ~\text{TT}}=-\frac{\Lambda }{\kappa }h_{ij}^{%
\text{TT}}  \label{T_ij (DE)^TT}
\end{equation}%
where $h_{ij}^{\text{TT}}\equiv h_{ij}-\tfrac{1}{3}H\delta _{ij}$. Comparing
the stress found in the \textit{material} constitutive equation $\left( \ref%
{T_ij (material)^TT}\right) $, and the \textit{gravitational} constitutive
equation $\left( \ref{T_ij (DE)^TT}\right) $, shows that there is
effectively a \textquotedblleft gravitational shear
modulus\textquotedblright\ given by%
\begin{equation}
\mu _{\text{G}}\equiv \frac{\Lambda }{\kappa }  \label{mu_G}
\end{equation}%
This quantity determines the shear \textquotedblleft
stiffness\textquotedblright\ of dark energy in response to a shear
gravitational wave strain field. Therefore $\left( \ref{T_ij (DE)^TT}\right) 
$ can now be written as%
\begin{equation}
T_{ij}^{\left( \text{DE}\right) ~\text{TT}}=-\mu _{\text{G}}h_{ij}^{\text{TT}%
}  \label{DE constitutive equation}
\end{equation}%

This equation was formally derived in \cite{FQMT15, Essay, Dissertation}
for describing the response of a superconductor to a gravitational wave. The
equation predicts that a shear stress, $T_{ij}^{TT}$, is caused in the
material due to the shear strain, $h_{ij}^{\text{TT}}$, of a gravitational
wave. In fact, from the point of view of \textit{electrical} constitutive
equations, $\mu _{\text{G}}$ is essentially the \textquotedblleft
conductivity\textquotedblright\ of the medium. In that context, dark energy
acts as a dissipationless \textquotedblleft medium\textquotedblright\
throughout the universe, where the \textquotedblleft
conductivity\textquotedblright\ is determined by $\Lambda $. The value can
be found from $\left( \ref{mu_G}\right) $ as%
\begin{equation}
\mu _{\text{G}}=\frac{c^{4}\Lambda }{8\pi G}\sim 10^{-10}\text{ J/m}^{3}
\label{mu_G (DE)}
\end{equation}%
This is the familiar value for the
cosmological energy density predicted by GR. By contrast, it is found in 
\cite{FQMT15, Essay, Dissertation} that the \textquotedblleft gravitational
shear modulus\textquotedblright\ of a superconductor is $\sim 10^{8}$ J/m$%
^{3}$. In that sense, dark energy can be modeled is an extremely weakly
interacting \textquotedblleft superconductor\textquotedblright\
dissipationless medium. However \cite{Beck, Matos} suggests that it might be possible to detect dark energy in an actual superconductor.\vspace*{0.3in}

\section{Screening of Newtonian gravity due to dark energy: the emergence of dark matter}

For a weak-field, $\left\vert h_{\mu \nu }\right\vert <<1$, it is shown in \cite{Dissertation} that
\begin{equation}
G_{00}=\tfrac{1}{2}\left( \partial _{i}\partial _{j}h_{ij}-\nabla ^{2}\delta
^{ij}h_{ij}\right) 
\end{equation}%
In the Newtonian limit, $h_{00}=-2\Phi /c^{2}$ and $h_{ij}=\delta _{ij}h_{00}
$, where $\Phi $ is the gravitational scalar potential.\footnote{%
With this definition, Newton's law of gravity is recovered by using $\vec{g}%
\equiv -\nabla \Phi $ and $T_{00}^{\text{matter}}=\rho _{\text{m}}c^{2}$,
where $\rho _{\text{m}}$ is the mass density, so that Einstein's equation
becomes $\nabla \cdot \vec{g}=-4\pi G\rho _{\text{m}}$.} Therefore $%
G_{00}=2\nabla ^{2}\Phi /c^{2}$. Letting $T_{\mu \nu }^{\text{matter}}=0$ in
the absence of normal matter sources and evaluating $\left( \ref{EFE}\right) 
$ for $\left( \mu ,\nu \right) =\left( 0,0\right) $ leads to%
\begin{equation}
\nabla ^{2}\Phi -\Lambda \Phi =c^{2}\Lambda /2
\end{equation}%
If we introduce a coordinate transformation to describe the scalar potential
due to dark energy as $\Phi _{\text{D}}\equiv \Phi +c^{2}/2$, then the static
limit in spherical coordinates gives%
\begin{equation}
\nabla ^{2}\Phi _{\text{D}}=\left( \frac{1}{r^{2}}\partial _{r}\right) \left(
r^{2}\partial _{r}\Phi _{\text{D}}\right) -\Lambda \Phi _{\text{D}}=0
\label{spherical}
\end{equation}%
For boundary conditions given by $\Phi_{\text{D}} \left( 0\right) =\Phi _{0}$ and $%
\underset{r\rightarrow \infty }{\lim }\Phi_{\text{D}} \left( r\right)=0$, the solution is%
\begin{equation}
\Phi_{\text{D}} \left( r\right) =-\frac{\mathcal{C}}{r}\Phi _{0}e^{-r/\lambda _{\text{G}%
}}  \label{solution}
\end{equation}%
where $\mathcal{C}$ is a constant determined by boundary
conditions, and%
\begin{equation}
\lambda _{\text{G}}\equiv \dfrac{1}{\sqrt{\Lambda }}\sim 10^{26}\text{ m}\sim \text{Gpc}
\label{lambda_G (Newtonian)},
\end{equation}%
which is the size of the observable universe and gives the characteristic length scale associated with the dark energy scalar potential. A different approach is used in \cite{Zhuk} which leads to a \textquotedblleft screening
length\textquotedblright\ given by $\lambda =\sqrt{\dfrac{c^{2}a^{2}H}{3}%
\bigintss \dfrac{da}{a^{3}H^{3}}}$. This length is stated as having a value of 2.57 Gpc which is the same order of magnitude as $\left( \ref {lambda_G (Newtonian)}\right)$. In what follows, we will show the emergence of the Yukawa gravitational potential in pure GR in the presence of dark energy, and we will explore some phenomenological aspects of this potential for the dynamics of galaxies and obtaining a cosmological model. 
\bigskip

\subsection{Recovering dark matter in galaxy scales}
Consider now a specific example of a given galaxy. In particular, we would like to see the effect of a massive graviton on the gravitational potential in the outer part of the galaxy where the Newtonian potential goes to zero. 
For reasons that we shall explain below, the dark matter potential is found by setting $\Phi_{\text{DM}}(r)=\Phi_{\text{D}}(r)$. Then $\left( \ref {spherical}\right)$ can be written as 
\begin{eqnarray}
    \frac{1}{r^2}\frac{\partial}{\partial r}\left( r^2 \frac{\partial \Phi_{\text{DM}}(r)}{\partial r}\right)=4 \pi G \rho_{\text{DM}},
\end{eqnarray}
where $\rho_{\text{DM}}$ is the effective energy density of dark matter which corresponds to $\Phi_{\text{DM}}(r)/\lambda^2_{\text{G}}=4 \pi G \rho_{\text{DM}}$. Note here that such matter should be viewed only as an apparent form of matter and not a real type of matter. Similar to  $\left( \ref{solution} \right)$, we have \begin{eqnarray}
    \Phi_{\text{DM}}(r)=-\frac{\mathcal{\mathcal{C}}}{r}e^{-r / \lambda_{\text{G}}}
\end{eqnarray}
This Yukawa-like relation describes the modification of the gravitational potential of a test particle with mass $m$ located at some distance $r$ from the galactic center, where for simplicity we may assume a constant source mass $M$. From dimensional units, it is therefore natural to assume that the constant $\mathcal{C}$ is linked to the mass $M$ and Newton's constant $G$. Let us take $\mathcal{C}=\alpha G M $, where $\alpha$ is a dimensionless quantity and encodes correlations between gravitons and the matter field, yielding 
\begin{eqnarray}
    \Phi_{\text{DM}}(r)=-\frac{\alpha G M }{r}e^{-\frac{r}{\lambda_{\text{G}}}}.
\end{eqnarray}
In the last equation, we claim that such a potential mimics the effect of dark matter. On the other hand, for the baryonic matter, we have the standard relation 
\begin{eqnarray}
    \frac{1}{r^2}\frac{\partial}{\partial r}\left( r^2 \frac{\partial \Phi_{\text{N}}(r)}{\partial r}\right)=0,
\end{eqnarray}
which has a solution in terms of the standard Newtonian gravitational potential, $\Phi_{\text{N}}(r)=-GM/r$. According to the last equation, the graviton has to be massless. In fact, as we will explain, describing galactic and cosmological dark matter phenomena requires the graviton mass to vary and depend on the environment. In our view, there are two possibilities: either a single tensor field can vary between massive and massless states depending on environmental factors such as matter densities, pressure, and temperature gradients, or there are two distinct tensor fields--one massless and one being always massive due to the uniform distribution of dark energy. This scenario is similar to the modified potential recently proposed in \cite{Abdelrahman:2024frk}. One can finally get the total contribution using the sum of both potentials. In doing so we get 
\begin{eqnarray}
    \Phi_{\rm total}(r)=-\frac{GM}{r}\left(1+\alpha e^{-\frac{r}{\lambda_{\text{G}}}}\right).
\end{eqnarray}
The last equation is an important result for the following reason: we obtained a Yukawa gravitational potential purely in the framework of GR when taking into account the presence of dark energy which naturally induces mass to the graviton. By contrast, Yukawa-like corrections to the gravitational potential in the weak field limit are typically obtained in extended theories of gravity, such as $f(R)$ models of gravity \cite{Capozziello:2007ms,Benisty:2023ofi}. It is interesting to point out that in many extended theories of gravity, in the Yukawa-like potential we have a contribution of a massless graviton as in General Relativity along with a massive scalar field that couples gravitationally to matter. In the present paper, we point out that the graviton effectively becomes massive due to the interaction with dark energy. Dark energy will be modeled as a superconductor in terms of some scalar field. This scalar field can, in principle, couple to matter as well. In this way we end up with a similar Yukawa-like potential. 
Using the above Yukawa potential, one can study the galactic dynamics by computing the force, $\vec{F}=-m\nabla \Phi_{\rm total}$, acting on a  test particle with mass $m$ having a circular speed which gives
\begin{equation}
F_r= - \frac{ G M m }{r^2} \left[1+\alpha\,\left(\frac{r+\lambda_{\text{G}}}{\lambda_{\text{G}}}\right)e^{-\frac{r}{\lambda_{\text{G}}}}\right].
\end{equation}
Alternatively, one can compute the circular velocity using $|F|=mv^2/r$ to obtain a MOND-like result:
\begin{equation}
\frac{v^2}{r} = \frac{G M }{r^2}+\sqrt{\left(\frac{GM}{r^2}\right)\left(\frac{G M (r+\lambda_{\text{G}})^2\alpha^2}{r^2 \lambda_{\text{G}}^2}e^{-\frac{2r}{\lambda_{\text{G}}}}\right)},
\end{equation}
This yields
\begin{eqnarray}
    a_{\rm total}=a_{\text{N}}+\sqrt{a_{\text{N}} a_{0}}=a_N+a_{\text{DM}},
\end{eqnarray}
where we defined $a_{\text{DM}}\equiv\sqrt{a_{\text{N}} a_{0}}$, along with
\begin{eqnarray}
    a_{\text{N}}\equiv\frac{G M }{r^2},
\end{eqnarray}
and 
\begin{eqnarray}
    a_{0}\equiv\frac{G M (r+\lambda_{\text{G}})^2\alpha^2}{r^2 \lambda_{\text{G}}^2}e^{-\frac{2r}{\lambda_{\text{G}}}}.
\end{eqnarray}
Furthermore, it is natural to assume that including matter (such as galaxies) into this dark energy model will introduce an interaction that will cause $\lambda_{\text{G}}$ and $m_{\text{G}}$ to vary on galactic length scales or environments. It follows that
\begin{equation}
\lambda _{\text{G}}(x)=\frac{\hslash }{m_{\text{G}}(x)c}\label{lambda_G (x)}
\end{equation}
Then differentiating $\left( \ref{lambda_G (x)}\right)$ and approximating $d \lambda_{\text{G}} / d m_{\text{G}} \simeq \Delta \lambda_{\text{G}} / \Delta m_{\text{G}}$ leads to
\begin{eqnarray}
  \Bigg|\frac{\Delta \lambda_{\text{G}}}{\lambda_{\text{G}}}\Bigg| \sim \Bigg|\frac{\Delta m_{\text{G}}}{m_{\text{G}}}\Bigg|.\label{frac change}
\end{eqnarray}
This relation gives the fractional change in the screening length scale (and the corresponding fractional change in the graviton mass scale) as a result of the interaction with galactic mass. In fact, recall that the Yukawa potential in $\left( \ref{solution}\right)$ was derived in the absence of normal matter. Therefore, any fractional change given by $\left( \ref{frac change}\right)$ would explain the observed effects of dark matter. On cosmological scales, $\lambda_{\text{G}}$ is on the order of Gpc, as shown by $\left( \ref {lambda_G (Newtonian)}\right)$, while on galactic scales, $\lambda_{\text{G}}$ is on the order of kpc \cite{Jusufi:2024rba,Jusufi:2024ifp}. This means for the outer part of the galaxy, $r \sim \lambda_G$, the acceleration is
\begin{eqnarray}
    a_{0}=\lim_{r \to \lambda_{\text{G}}}\frac{G M (r+\lambda_{\text{G}})^2\alpha^2}{r^2 \lambda_{\text{G}}^2}e^{-\frac{2r}{\lambda_{\text{G}}}} = {\rm constant},
\end{eqnarray}
which can describe the flat rotating curves of galaxies. In the above discussion, $M$ is treated as a constant (which is a good approximation if one studies the motion of a test particle in the outer part of the galaxy), however, in general, one can take a specific function of mass distribution $M(r)$. In that case one has the Newtonian force  $F(r)=|-m\nabla \Phi_{\rm{total}}(r)|=m v^2(r)/r$ along with the relation for the circular velocity, $v^2(r)=r \nabla\Phi_{\rm{total}}(r)$, and $v^2_{\text{N}}(r)=r \nabla\Phi_{\text{N}}(r)=GM(r)/r$, respectively. Using the Yukawa potential leads to
\begin{equation}
    r \nabla\Phi_{\rm{tot}}(r)=r \nabla\Phi_{\text{N}}(r)\left(1+ \alpha e^{-\frac{r}{\lambda_{\text{G}}}}\right)-\frac{\alpha  r \Phi_{\text{N}}(r) e^{-\frac{r}{\lambda_{\text{G}}}}}{\lambda_{\text{G}}},
\end{equation}
which becomes 
\begin{eqnarray}
    v^2(r)=v^2_{\text{N}}(r)+v^2_{\text{DM}}(r)
    \label{v^2},
\end{eqnarray}
where we have used $\Phi(r)=-GM(r)/r$ and defined 
\begin{eqnarray}
    v^2_{\text{DM}}(r)\equiv\frac{G M_{\text{DM}}(r)}{r}=v^2_{\text{N}}(r) \alpha \left( \frac{r+\lambda_{\text{G}}}{\lambda_{\text{G}}} \right)e^{-\frac{r}{\lambda_{\text{G}}}}.
\end{eqnarray}
Then the dark matter mass is obtained from 
\begin{eqnarray}
    M_{\text{DM}}(r)=\alpha M(r) \left( \frac{r+\lambda_{\text{G}}}{\lambda_{\text{G}}} \right)e^{-\frac{r}{\lambda_{\text{G}}}}.
\end{eqnarray}
In the outer part of the galaxy, we can take $r \sim \lambda_G$ and, in doing so, we can use the fact that the Newtonian term for the velocity vanishes, i.e., $v^2_{\text{N}}(r)=GM(r)/r \to 0$ for large distances, hence by neglecting the first term in $\left( \ref{v^2}\right)$ we get for the velocity only the dark matter contribution
\begin{equation}
    v^2(r) =  \lim_{r \to \lambda_{\text{G}}}  \frac{\alpha G M}{r}  \left( \frac{r+\lambda_{\text{G}}}{\lambda_{\text{G}}} \right)e^{-\frac{r}{\lambda_{\text{G}}}}=\rm{constant}.
\end{equation}
In this equation, mass can now be viewed as a constant term and the total result is again a constant. This equation is therefore in agreement with the observations of rotating flat curves. Recently, a Yukawa-like gravitational potential in extended theories of gravity (such as the $f(R)$ gravity) was tested for the Milky Way and M31 galactic scales, and it was shown that indeed such a potential can explain the galactic dynamics and rotating curves \cite{DAgostino:2024ojs}. The important finding in the present paper is that a Yukawa gravitational potential naturally emerges from pure GR in the presence of dark energy. One can therefore explain the extra force that accounts for dark matter in terms of a long-range force which is a consequence of the graviton having mass that is induced by dark energy, and the correlations between gravitons and matter fields. This shows that the fundamental quantity in our model is dark energy - which is modeled as a superconductor, while dark matter is only an emergent effect. 

\subsection{Recovering dark matter in cosmological scales}
The Yukawa gravitational potential significantly influences cosmological theories, particularly by considering a spatially homogeneous and isotropic background spacetime, as described by the Friedmann-Robertson-Walker (FRW) metric
\begin{equation}
ds^2=-dt^2+a^2\left[\frac{dr^2}{1-kr^2}+r^2(d\theta^2+\sin^2\theta
d\phi^2)\right]
\end{equation}
where $a(t)$ is the cosmological scale
factor. We can extend the analysis by introducing the transformation $R=a(t)r$, alongside the two-dimensional metric $h_{\mu \nu}$. In this context, the parameter $k$ denotes the curvature of space, which we set to $k=0$ to describe a spatially flat universe. In such a universe, the presence of a dynamic apparent horizon can be determined through the following calculation:
$h^{\mu
\nu}(\partial_{\mu}R)\,(\partial_{\nu}R)=0$, yielding the apparent horizon radius 
\begin{equation}
\label{radius}
 R=ar=H^{-1}.
\end{equation}
In addition one can assume the presence of a perfect
fluid described by the stress-energy tensor
\begin{equation}\label{T}
T_{\mu\nu}=(\rho+p)u_{\mu}u_{\nu}+pg_{\mu\nu},
\end{equation}
along with the continuity equation given by
\begin{equation}\label{Cont}
\dot{\rho}+3H(\rho+p)=0,
\end{equation}
with $H=\dot{a}/a$ being the Hubble parameter. Consider a compact spatial region with a compact boundary that is a sphere of radius $R= a(t)r$, where $r$ is a dimensionless quantity. Also assume that the Yukawa potential evolves according to 
\begin{eqnarray}
    \Phi_{\rm total}(R)=-\frac{G M }{R}\left(1+\alpha e^{-\frac{R}{\lambda_{\text{G}}}}\right).
\end{eqnarray}
This leads to a Newtonian force for the test particle $m$ near the surface given by
\begin{equation}
F=m\ddot{a}r=-\frac{G m \left(M(R)+M_{\text{DM}}(R)\right)}{R^2},
\end{equation}
where 
\begin{eqnarray}
   M_{\text{DM}}(R)= \alpha\,M(R)\left(\frac{R+\lambda_{\text{G}}}{\lambda_{\text{G}}}\right)e^{-\frac{R}{\lambda_{\text{G}}}}.
\end{eqnarray}
The transition from a dynamical equation in Newtonian-like cosmology to the fully relativistic modified Friedmann equations of the FRW universe in GR, is possible if we use the active gravitational mass, denoted by $\mathcal{M}$, rather than the total mass $M$. This active gravitational mass (Komar mass) is defined as
\begin{equation}
\mathcal M =2
\int_V{dV\left(T_{\mu\nu}-\frac{1}{2}Tg_{\mu\nu}\right)u^{\mu}u^{\nu}}.
\end{equation}
From here, one can show
\begin{equation}
\mathcal M =\sum_i(\rho_i+3p_i)\frac{4\pi G}{3}a^3 r^3,
\end{equation}
where we have assumed several matter fluids with a constant equation of state parameters $\omega_i$ satisfying $\dot{\rho}_i+3H(1+ \omega_i) \rho_i=0.$ This leads to the following relation  \cite{Jusufi:2023xoa}
\begin{equation}
\frac{\ddot{a}}{a}=- \left(\frac{4 \pi G }{3}\right)\sum_i \left(\rho_i+3p_i\right) \left[1+\alpha\,\left(\frac{R+\lambda_{\text{G}}}{\lambda_{\text{G}}}\right)e^{-\frac{R}{\lambda_{\text{G}}}}\right], 
\end{equation}
Expanding upon the given expression for densities, $\rho_i=\rho_{i0} a^{-3(1+\omega_i)}$, multiplying both sides by $2\dot{a}a$, and performing some algebraic manipulation leads to
\begin{align}
 \dot{a}^2+C = &   \frac{8\pi G}{3} \int \left[1+\alpha (1+x) e^{-x}\right] \frac{d \left(\sum_i \rho_{i 0} a^{-1-3\omega_i}\right)}{da} da,
\end{align}
where we define $x \equiv R/\lambda_G$, with $r$ nearly constant and $C$ as an integration constant. We make the assumption that the quantity $x$ is a small number since $\lambda_{\text{G}}$ is of the same order as $R(a)$. Further, we utilize the approximation $1+\alpha(1+x)e^{-x}\simeq 1+\alpha-\alpha R^2/2\lambda_{\text{G}}^2$ to derive
\begin{align}
H^2= & \frac{8\pi G_{\rm eff}
}{3}\sum_i \rho_i +\frac{4 \pi G_{\rm eff}}{3}R^2 \sum_{i}\Gamma (\omega_i)\rho_i, 
\end{align}
where $ G_{\rm {eff}}\equiv G (1+\alpha)$, and \begin{equation}
\Gamma \left( \omega _{i}\right) \equiv \left. \dfrac{\alpha \left(
1+3\omega _{i}\right) }{\lambda _{\text{G}}^{2}\left( 1+\alpha \right)
\left( 1-3\omega _{i}\right) }\right\vert _{\omega _{i}=0}
\end{equation}%
Consider a universe comprised solely of baryonic matter $(\omega=0)$, radiation $(\omega=1/3)$, and dark energy $(\omega=-1)$, where $\Gamma$ influences the late-time universe dynamics. Notably, the last equation reveals a singularity for radiation $(\omega=1/3)$, indicating a phase transition from a radiation-dominated to a matter-dominated state in the early Universe \cite{Jusufi:2023xoa}. In a radiation-dominated universe, this suggests that $\alpha=0$ and therefore $\Gamma=0$, and thus, no singularity arises, implying the significance of $\alpha$ only post-transition. Using $\rho_{\rm crit}=\frac{3}{8 \pi G}H_0^2$ and solving for $E(z)=H/H_0$ in terms of the redshift $a^{-1}=1/(1+z)$, yields \cite{Jusufi:2023xoa,Jusufi:2024ifp}
\begin{align}\label{eq43}
   E(z)&=\frac{(1+\alpha)}{2}\,\sum_i\Omega_i \notag \\
   &\pm  \frac{\sqrt{(\sum_i\Omega_i)^2 (1+\alpha)^2+2 \Gamma (\omega_i) \Omega_i (1+\alpha)/H_0^2}}{2},
\end{align}

As was argued in \cite{Jusufi:2023xoa}, the physical interpretation of the term $
    2  \Gamma (\omega_i) \Omega_i (1+\alpha)/H_0^2 $ is closely linked to the presence of dark matter which appears as an apparent effect in Yukawa cosmology. When considering  $\omega_i=0$, i.e., the effect of cold dark matter, we have
    \begin{equation}
   \frac{\Omega^2_{DM}(1+\alpha)^2}{{(1+z)^3}}\equiv \frac{2 \Gamma \Omega_i(1+\alpha)}{H_0^2}\Bigg|_{\omega_i=0}.
\end{equation}
Specifically, it has been shown that the density parameter for dark matter can be related to baryonic matter as \cite{Jusufi:2023xoa}
\begin{equation}\label{eqDM}
    \Omega_{DM}= \frac{c \sqrt{2 \alpha \Omega_{\text{B,}0}}}{\lambda_G H_0\,(1+\alpha)} 
\,{(1+z)^{3}},
\end{equation}
where we have introduced the constant $c$. The subscript \textquotedblleft 0\textquotedblright\ denotes quantities evaluated at present, specifically at $z=0$, implying dark matter's interpretation as a consequence of the modified Newtonian law characterized by $\alpha$ and $\Omega_{\text{B}}$. We can assume the constant $c$ satisfies the following definition \cite{Jusufi:2023xoa}
\begin{equation}
    \Omega_{\Lambda,0}\equiv \frac{c^2\,\alpha}{\lambda_{\text{G}}^2 H^2_0 (1+\alpha)^2}. \label{eq:(20)}
\end{equation}
Then an expression can be derived that establishes a relation between baryonic matter, effective dark matter, and dark energy as
\begin{equation}\label{DMCC}
   \Omega_{\text{DM}}(z)= \sqrt{2\,\Omega_{\text{B},0}  \Omega_{\Lambda,0}}{(1+z)^3}\,.
\end{equation}
Considering as a physical solution only the one with the positive sign, we have \cite{Jusufi:2024ifp}
\begin{align}\label{eq43B}
  E^2(z)&=\frac{(1+\alpha)}{2}\,\left(\Omega_{\text{B},0} 
(1+z)^{3}+\Omega_{\Lambda,0}\right)\notag \\
   &+\frac{(1+\alpha)}{2}\sqrt{\left(\Omega_{\text{B},0} 
(1+z)^{3}+\Omega_{\Lambda,0}\right)^2+\frac{\Omega^2_{\text{DM}}(z)}{{(1+z)^3}}}.
\end{align}
which describes the Yukawa-modified cosmological model and was explored recently in \cite{Jusufi:2024ifp,Gonzalez:2023rsd}.

\section{Graviton mass from the Cosmological Constant}

For weak gravitational fields, it has been shown in \cite{FlanaganandHughes} that the only propagating degrees of freedom in linearized GR are
the transverse-traceless spatial perturbations, $h_{ij}^{\text{TT}}$, where $%
\partial _{i}h_{ij}^{\text{TT}}=0$ and $\delta ^{ij}h_{ij}^{\text{TT}}=0$.
Then the linearized Einstein field equation gives%
\begin{equation}
\square h_{ij}^{\text{TT}}=-2\kappa T_{ij}^{\text{TT}}  \label{GR wave}
\end{equation}%
Substituting $\left( %
\ref{T_ij (DE)^TT}\right) $ into $\left( \ref{GR wave}\right) $, and letting $T_{\mu
\nu }^{\text{matter}}=0$ in the absence
of normal matter sources, leads to%
\begin{equation}
\square h_{ij}^{\text{TT}}-k_{\text{G}}^{2}h_{ij}^{\text{TT}}=0
\label{wave equ with constit equ}
\end{equation}%
where $k_{\text{G}}^{2}\equiv 2\Lambda $. Since $\left( \ref{wave equ with
constit equ}\right) $ has the same form as $\left( \ref{K-G (flat)}\right) $
and $\left( \ref{EM wave}\right) $, then the second term in $\left( \ref%
{wave equ with constit equ}\right) $ can be interpreted as a
\textquotedblleft mass term\textquotedblright\ which implies that the
gravitational wave can be viewed as a massive tensor field due to the
interaction with dark energy. Using $\left( \ref{mass scale}\right) $, the
graviton mass is $m_{\text{G}}=k_{\text{G}}\hslash /c$ which leads to%
\footnote{%
This is consistent with \cite{Visser} which gives an upper bound of $m_{%
\text{G}}<2\times 10^{-29}$ eV $\approx 4$ $\times 10^{-65}$ kg. Also see 
\cite{Will} for various limits on graviton mass values.}%
\begin{equation}
m_{\text{G}}=\frac{\hslash }{c}\sqrt{2\Lambda }\sim 10^{-68}\text{ kg}
\label{m_graviton}
\end{equation}%
Therefore, the graviton is massive due to the interaction of gravitational waves
with dark energy acting as a \textquotedblleft medium\textquotedblright\
similar to a superconductor. Note that the photon is massless when it is in
vacuum, but is effectively massive in a superconducting medium. However, for
the case of the graviton, it must \textit{always} be massive because dark
energy is a \textquotedblleft medium\textquotedblright\ that pervades the
entire universe.\bigskip

Also using $\lambda _{\text{G}}^{\prime }\equiv 1/k_{\text{G}}$, we find that the
length scale associated with the Compton wave number is\footnote{%
This is again consistent with \cite{Visser} which gives a lower bound of $%
\lambda _{\text{G}}^{\prime }>6\times 10^{22}$ m.}%
\begin{equation}
\lambda _{\text{G}}^{\prime }=\frac{\hslash }{m_{\text{G}}c}=\dfrac{1}{\sqrt{%
2\Lambda }}  \label{lambda_G (waves)}
\end{equation}%
which is similar to $\left( \ref{lambda_G (Newtonian)}\right) $. This result gives an effective screening length scale for gravitational waves propagating through the universe. 
\bigskip 

It is stated in Section 8.4 of \cite{Ryder} that \textquotedblleft Another
way of stating the Meissner effect is to say that the photons are
effectively massive . . .\textquotedblright\ Therefore the fact that the
graviton becomes massive (as described above)\ implies that dark energy
produces a gravitational Meissner-like effect. In fact, just as the Meissner
effect can cause a repulsive force on a magnet, so also dark energy causes
an effective \textquotedblleft repulsion\textquotedblright\ on space-time
itself which causes the expansion of the universe. The gravitational
Meissner-like effect will be investigated in a later section.\vspace*{0.3in}

\section{Gravitational wave plasma frequency and penetration depth}

Now consider a monochromatic plane wave given by $h_{ij}^{\text{TT}}\left( 
\vec{x},t\right) =A_{ij}^{\text{TT}}e^{i\left( \vec{k}\cdot \vec{x}-\omega
t\right) }$, where $A_{ij}^{\text{TT}}$ is a constant amplitude tensor.
Using this wave solution in $\left( \ref{wave equ with constit equ}\right) $
leads to a dispersion relation given by%
\begin{equation}
k^{2}=\dfrac{\omega ^{2}}{c^{2}}\left( 1-\dfrac{2c^{2}\Lambda }{\omega ^{2}}%
\right)  \label{dispersion of GR waves in SC '}
\end{equation}%
It is pointed out in \cite{Press} that $\left( \ref{dispersion of GR waves
in SC '}\right) $ resembles the electromagnetic equation for a dense plasma.
Then the universe has a \textquotedblleft gravitational plasma
frequency\textquotedblright\ which can be defined as
\begin{equation}
\omega _{\text{G}}\equiv c\sqrt{2\Lambda }\sim 10^{-18}~\text{rad/s}
\label{w_G}
\end{equation}%
The standard meaning of a plasma frequency is that a material becomes
effectively transparent for frequencies above the plasma frequency. The vast
majority of gravitational wave frequencies tend to be above this, even for
ultra-low gravitational wave research \cite{Moore,Agazie}. However, $%
\left( \ref{w_G}\right) $ is the lower bound associated with cosmological
events such as quantum fluctuations in the early epochs of the universe
(which have been amplified by inflation), first-order phase transitions, and isolated loops of cosmic strings that decay through gravitational waves \cite{GravCMB}. Such gravitational waves would therefore experience attenuation due to dark energy while propagating through the universe. In fact, a penetration depth (or screening length scale) is calculated next.\\

We can  define a complex wave number as $k=K+i\alpha $, where $K$ and $%
\alpha $ are real quantities, and insert this into the plane wave solution.
Separating the real and imaginary parts of the phase gives
\begin{equation}
h_{ij}^{\text{TT}}\left( \vec{x},t\right) =A_{ij}^{\text{TT}}e^{-\alpha
x}e^{i\left( \vec{K}\cdot \vec{x}-\omega t\right) }
\end{equation}%
Here we see that the wave falls off exponentially with distance, where $%
\alpha $ is the exponential decay factor. The square of the wave number is $%
k^{2}=K^{2}-\alpha ^{2}+2iK\alpha $. Since $k^{2}$ in $\left( \ref%
{dispersion of GR waves in SC '}\right) $ is only real, then we must have
either $K=0$ or $\alpha =0$. For $K=0$, we use $\left( \ref{dispersion of GR
waves in SC '}\right) $ to solve for $\alpha $ and define a
frequency-dependent penetration depth as $\delta _{\text{G}}\equiv 1/\alpha $%
. This leads to solution given by $h_{ij}^{\text{TT}}\left( \vec{x},t\right)
=A_{ij}^{\text{TT}}e^{-x/\delta _{\text{G}}}e^{-i\omega t}$, where%
\begin{equation}
\delta _{\text{G}}^{2}=\dfrac{c^{2}}{2c^{2}\Lambda -\omega ^{2}}
\label{delta_G from CC}
\end{equation}%
An exponential decay solution implies that as gravitational waves propagate
through the universe, they are attenuated due to the presence of dark energy. In the DC limit $\left( \omega =0\right) $, the penetration depth (or screening length scale) in 
$\left( \ref{delta_G from CC}\right) $ has an upper bound of $\delta _{\text{%
G}}=1/\sqrt{2\Lambda }$ which is consistent with $\left( \ref{lambda_G
(waves)}\right) $.\vspace*{0.3in}

\section{Gravitational wave index of refraction, phase velocity, and group velocity}

Rearranging $\left( \ref{dispersion of GR waves in SC '}\right) $ and using $\left( \ref{w_G}\right) $ leads to
\begin{equation}
\dfrac{\omega ^{2}}{k^{2}}=\dfrac{c^{2}}{1-\omega _{\text{G}}^{2}/\omega ^{2}%
}  \label{omega^2 / k^2}
\end{equation}%
Solving for $v_{\text{p}}=\omega /k$ gives the phase velocity as%
\begin{equation}
v_{\text{p}}=\dfrac{c}{\sqrt{1-\omega _{\text{G}}^{2}/\omega ^{2}}}
\label{v_phase}
\end{equation}%
Using $\left( \ref{w_G}\right) $ makes $\left( \ref{dispersion of GR waves
in SC '}\right) $ become%
\begin{equation}
k^{2}=\dfrac{\omega ^{2}}{c^{2}}\left( 1-\dfrac{\omega_{\text{G}}^{2}}{%
\omega ^{2}}\right)  \label{k in terms of omega_G '}
\end{equation}%
It is clear from $\left( \ref{k in terms of omega_G '}\right) $ that
transmission no longer occurs when $\omega \leq \omega _{\text{G}}$ since $k$
becomes imaginary. Also, if the phase speed is $v_{\text{p}}=\omega /k$ , and the
\textquotedblleft gravitational index of refraction\textquotedblright\ is $%
n_{\text{G}}=v/c$, then $\left( \ref{k in terms of omega_G '}\right) $ can
be written as%
\begin{equation}
k^{2}=\dfrac{\omega ^{2}}{c^{2}}n_{\text{G}}^{2}\left( \omega \right)
\end{equation}%
Matching this result with $\left( \ref{k in terms of omega_G '}\right) $
implies that the gravitational index of refraction due to dark energy is%
\begin{equation}
n_{\text{G}}\left( \omega \right) \equiv \sqrt{1-\omega _{\text{G}%
}^{2}/\omega ^{2}}  \label{n_G}
\end{equation}%
Notice that in the absence of dark energy, $\Lambda =0$, so $\omega _{\text{G%
}}=0$ and therefore $n_{\text{G}}=1$. This corresponds to the standard
result that $v_{\text{p}}=c$ for gravitational waves in vacuum. However,
since the presence of dark energy implies $\Lambda \neq 0$, then we may
consider cases which would lead to gravitational waves propagating at a speed other than $c$.

For example, LIGO operates in the frequency range of 10 Hz to 10
kHz \cite{LIGO}. This corresponds to a maximum value of $n_{\text{G}%
}\approx 1-10^{-38}$ which is totally negligible. However, for frequencies
as low as $\omega \sim 10^{-17}~$rad/s, then $n_{\text{G}}\sim 1-10^{-5}$.
If such sources are in the furthest observable reaches of the universe $%
\left( L\sim 10^{26}\text{ m away}\right) $, the time it would take
gravitational waves to reach earth would be $t=L/v_{\text{p}}=Ln_{\text{G}%
}/c\sim \left( 1-10^{-5}\right) T_{\text{0}}$, where $T_{\text{0}}\sim
10^{17}$ s is the age of the observable universe determined by using light.
This means that gravitational waves would arrive early (compared to light)\
by $\left(10^{-5}\right)T_{\text{0}}\sim 10^{12}$ s $\sim 10^{4}$ years.  In other words, if the age of the observable universe was determined by low-frequency
gravitational waves (rather than light), it would be thought to be younger
by $10^{4}$ years. The current uncertainty in the age of the universe is $\sim 10^{7}$ years \cite{Planck_collab}. Therefore, this
discrepancy would not be noticed. However, there are efforts to measure the
Hubble constant using gravitational waves, and to compare the result to
standard measurements involving light \cite{LIGO}. There is also a comprehensive study in \cite{Grav_wave_speeds} showing various possibilities for the speed of gravitational waves compared to the speed of light.\bigskip 

For frequencies much greater
than the gravitational plasma frequency, $\omega ^{2}>>\omega _{\text{G}}^{2}
$, we can use a binomial expansion (to first order) to write $\left( \ref{v_phase}%
\right) $ as $v_{\text{p}}\approx c\left[ 1+\omega _{\text{G}}^{2}/\left(
2\omega ^{2}\right) \right] $. This implies that the phase velocity will be
superluminal. We can also see this from the gravitational index of
refraction given in $\left( \ref{n_G}\right) $ as $n_{\text{G}}\left( \omega
\right) =\sqrt{1-\omega _{\text{G}}^{2}/\omega ^{2}}$. For $\omega
^{2}>>\omega _{\text{G}}^{2}$, we have $n_{\text{G}}\lesssim 1$. Then $v_{%
\text{p}}=c/n_{G}$ implies that $v_{\text{p}}\gtrsim c$. For frequencies just above the gravitational plasma frequency, $\omega
^{2}\gtrsim \omega _{\text{G}}^{2}$, we find that $v_{\text{p}}$ in $\left( %
\ref{v_phase}\right) $ becomes arbitrarily large. As $\omega $ approaches $%
\omega _{\text{G}}$, then $v_{\text{p}}$ diverges to infinity This
corresponds to $n_{\text{G}}$ going to zero so that $v=c/n_{\text{G}}$
becomes infinite.\bigskip

Lastly, when $\omega <\omega _{\text{G}}$, then $v_{\text{p}}$ and $n_{\text{%
G}}$ both become imaginary. This implies a complete expulsion of the wave
such that there is no phase velocity of the wave. Since the medium is
completely dissipationless, then there can be no absorption of the wave at
all. Rather, there must be a perfect external reflection of the wave.\bigskip

Next, we consider the \textit{group} velocity of the wave. Returning to $%
\left( \ref{omega^2 / k^2}\right) $ and expressing $\omega ^{2}$ in terms of 
$k^{2}$ gives%
\begin{equation}
\omega ^{2}=c^{2}k^{2}+\omega _{\text{G}}^{2}
\label{omega^2 in terms of k^2}
\end{equation}%
Taking the derivative with respect to $k$, and solving for $v_{\text{g}}=%
\dfrac{d\omega }{dk}$ gives $v_{\text{g}}=c^{2}k/\omega $. Solving $\left( %
\ref{omega^2 in terms of k^2}\right) $ for $k$ yields $k=\dfrac{\omega }{c}%
\sqrt{1-\omega _{\text{G}}^{2}/\omega ^{2}}$. Then the group velocity becomes%
\footnote{%
The expression in $\left( \ref{v_group}\right) $ is consistent with \cite%
{Visser} which has $v_{\text{g}}=c\sqrt{1-\lambda ^{2}/\lambda _{\text{G}%
}^{2}}$.}%
\begin{equation}
v_{\text{g}}=c\sqrt{1-\omega _{\text{G}}^{2}/\omega ^{2}}  \label{v_group}
\end{equation}%
For frequencies much greater than the gravitational plasma frequency, $%
\omega ^{2}>>\omega _{\text{G}}^{2}$, then we have $\omega _{\text{G}%
}^{2}/\omega ^{2}<<1$ which means we can use a binomial expansion to first
order to obtain $v_{\text{g}}\approx c\left[ 1-\omega _{\text{G}}^{2}/\left(
2\omega ^{2}\right) \right] $. This implies that for an arbitrarily large $%
\omega $, we can make $v_{\text{g}}$ arbitrarily close to $c$. This would
describe a wave that is almost completely unaffected by a medium and
therefore propagates through it at almost the same speed it has in
vacuum.\bigskip

Notice that $v_{\text{g}}$ is always subluminal. In fact, $\left( \ref%
{v_group}\right) $ can be written as $v_{\text{g}}^{2}=c^{2}-c^{2}\omega _{%
\text{g}}^{2}/\omega ^{2}$ which means that $v_{\text{g}}^{2}$ always
remains less than $c^{2}$ by an amount $c^{2}\omega _{\text{g}}^{2}/\omega
^{2}$. As $\omega $ decreases, $v_{\text{g}}$ in $\left( \ref{v_group}%
\right) $ will decrease until it vanishes when $\omega =\omega _{\text{G}}$.
For $\omega ^{2}<\omega _{\text{G}}^{2}$, then $v_{\text{g}}$ becomes
imaginary. Once again, this implies a complete expulsion of the wave at
these frequencies such that there is no group velocity of the wave. These
results collectively show that $v_{\text{g}}$ is always subluminal which is
expected since $v_{\text{g}}$ is the rate at which energy (and information)\
can be transported in the medium.\vspace*{0.3in}

\section{Gravitational wave impedance}

Recall that in electromagnetism, the impedance is found from the ratio of $%
\vec{E}$ and $\vec{H}$, where $\vec{H}=\vec{B}/\mu $ and $\vec{B}=\vec{E}/v_{%
\text{p}}$. This gives%
\begin{equation}
Z=\dfrac{\vec{E}}{\vec{H}}=\dfrac{\vec{E}}{\vec{B}/\mu }=\dfrac{\vec{E}}{%
\vec{E}/\left( v_{\text{p}}\mu \right) }=v_{\text{p}}\mu   \label{Z}
\end{equation}%
In vacuum, $v_{\text{p}}=c$ which leads to the standard result of $%
Z_{0}=c\mu _{0}=\sqrt{\mu _{0}/\varepsilon _{0}}$.
Similarly, for gravitational waves in vacuum, we can define a \textquotedblleft
gravitational permittivity\textquotedblright\ $\varepsilon _{\text{G}}\equiv
\left( 4\pi G\right) ^{-1}$, and a \textquotedblleft gravitational
permeability\textquotedblright\ $\mu _{\text{G}\left( \text{vac}\right)
}\equiv 4\pi G/c^{2}$, to obtain%
\begin{equation}
Z_{\text{G}\left( \text{vac}\right) }=\sqrt{\dfrac{\mu _{\text{G}\left( 
\text{vac}\right) }}{\varepsilon _{\text{G}}}}=\dfrac{4\pi G}{c}\approx
2.8\times 10^{-18}\text{ m}^{2}/\left( \text{kg}\cdot \text{s}\right) 
\label{Z_G^(vac)}
\end{equation}%
This result plays the same role as the electromagnetic wave impedance in
vacuum, namely, it characterizes the intrinsic impedance associated with a
wave propagating through empty space. For gravitational waves in a medium,
we must use a ratio of fields analogous to $\vec{E}$ and $\vec{H}$. As shown
in \cite{FQMT15}, \cite{Dissertation} and \cite{Barnett}, we can define%
\begin{equation}
\mathcal{E}_{ij}^{TT}\equiv -\partial _{t}h_{ij}^{\text{TT}}\qquad \text{and}%
\qquad \mathcal{B}_{ij}\equiv \varepsilon _{ikl}\partial _{k}h_{lj}^{\text{TT%
}}  \label{electric-like and magnetic-like tensor fields '}
\end{equation}%
which are electric-like and magnetic-like tensor fields for gravitational
waves analogous to $\vec{E}=-\partial _{t}\vec{A}$ and $\vec{B}=\nabla
\times \vec{A}$, respectively. We can also define an auxiliary magnetic-like
gravitational wave tensor field as%
\begin{equation}
\mathcal{H}_{ij}^{\text{TT}}=\mathcal{B}_{ij}^{\text{TT}}/\mu _{G}
\end{equation}%
Then by analogy with electromagnetism, we can define a gravitational wave
impedance as\footnote{%
Here we let $h_{ij}^{\text{TT}}$ be a plane-fronted monochromatic wave, $%
h_{ij}^{\text{TT}}=A_{ij}^{\text{TT}}e^{i\left( \vec{k}\cdot \vec{x}-\omega
t\right) }$, so that $\partial _{t}h_{ij}^{\text{TT}}=-i\omega h_{ij}^{\text{%
TT}}$, as well as $\partial _{i}h_{ij}^{\text{TT}}=ikh_{ij}^{\text{TT}}$. We
also use $k=\omega /v_{\text{p}}$.}%
\begin{equation}
Z_{\text{G}}\equiv \dfrac{\mathcal{E}_{ij}^{\text{TT}}}{\mathcal{H}_{ij}^{%
\text{TT}}}=\dfrac{-\partial _{t}h_{ij}^{\text{TT}}}{\mathcal{\partial }%
_{k}h_{ij}^{\text{TT}}/\mu _{\text{G}}}=\dfrac{\omega \mu _{\text{G}}}{k}=v_{%
\text{p}}\mu _{\text{G}}  \label{Z_GR wave}
\end{equation}%
This result is directly analogous to the case for electromagnetism in $%
\left( \ref{Z}\right) $. Notice that setting $v_{\text{p}}\mu _{\text{G}%
}=c\mu _{\text{G}\left( \text{vac}\right) }$ gives the vacuum result in $%
\left( \ref{Z_G^(vac)}\right) $. On the other hand, using $\mu _{\text{G}%
}=4\pi G/v^{2}$ and $v=c/n_{\text{G}}$ gives $Z_{\text{G}}=4\pi Gn_{\text{G}%
}/c=Z_{\text{G}\left( \text{vac}\right) }n_{\text{G}}$. This shows that a larger
gravitational index of refraction implies that the medium is more optically
dense to gravitational waves, and hence the gravitational wave impedance
will be larger. Using $\left( \ref{w_G}\right) $ and $\left( \ref{n_G}%
\right) $ gives the gravitational wave impedance due to dark energy as%
\begin{equation}
Z_{\text{G}}=\dfrac{Z_{\text{G}\left( \text{vac}\right) }}{\sqrt{%
1-2c^{2}\Lambda /\omega ^{2}}}
\end{equation}%
This implies that if $\omega >>2c^{2}\Lambda $, then $Z_{\text{G}}$ reduces
to the vacuum result in $\left( \ref{Z_G^(vac)}\right) $ which means the
dark energy medium does not attenuate the gravitational wave significantly.
In order to have significant attenuation, we must have $\omega \lesssim c%
\sqrt{2\Lambda }\sim 10^{-18}~$rad/s. Such frequencies are associated with
the particular cosmological events that were described in Section V.\vspace*{%
0.3in}

\section{Gravitational Meissner-like effect due to dark energy}

First, we briefly review the standard Meissner effect for electromagnetism in
a superconductor. Using the London constitutive equation, $\vec{J}=-\Lambda
_{\text{L}}\vec{A}$, and taking temporal and spatial derivatives leads to
constitutive equations involving the electric and magnetic fields within a
superconductor given as, respectively,%
\begin{equation}
\partial _{t}\vec{J}=\Lambda _{\text{L}}\vec{E}\qquad \text{and}\qquad
\nabla \times \vec{J}=-\Lambda _{\text{L}}\vec{B}  \label{London equations}
\end{equation}%
For a sinusoidal current density, we have $\partial _{t}J\propto \omega \vec{%
J}$. Therefore, in the DC limit $\left( \omega \rightarrow 0\right) $, the
first equation in $\left( \ref{London equations}\right) $ requires that $%
\vec{E}=0$. This implies that in the DC limit, the electric field vanishes
completely throughout the entire superconductor and only a magnetic field
remains within the London penetration depth of the superconductor. The
magnetic field drives the supercurrents by the second constitutive equation
in $\left( \ref{London equations}\right) $. Furthermore, taking the curl of
Ampere's law, $\nabla \times \left( \nabla \times \vec{B}\right) =\mu
_{0}\nabla \times \vec{J}$, and using $\vec{J}=-\Lambda _{\text{L}}\vec{A}$
leads to a Yukawa-like equation for the magnetic field given as%
\begin{equation}
\nabla ^{2}\vec{B}-\dfrac{1}{\lambda _{\text{L}}^{2}}\vec{B}=0
\label{Yukawa}
\end{equation}%
For boundary conditions given by $\vec{B}\left( 0\right) =\vec{B}_{0}$ and $%
\underset{\vec{x}\rightarrow \infty }{\lim }\vec{B}\left( x\right)= 0$, the solution is $\vec{B}\left( x\right) =\vec{B}%
_{0}e^{-x/\lambda _{\text{L}}}$, where $x$ is the distance into the surface
of the superconductor, and $\lambda _{\text{L}}$ is the London penetration
depth found in $\left( \ref{lamda_L}\right) $. This result demonstrates
that the magnetic field is expelled from the interior of the superconductor
which is referred to as the Meissner effect.\bigskip

An analogous approach can be used to demonstrate a Meissner-like effect
for the DC limit of gravitational waves due to dark energy. Taking
temporal and spatial derivatives of $\left( \ref{DE constitutive equation}%
\right) $ and using $\left( \ref{electric-like and magnetic-like tensor
fields '}\right) $ leads to the following constitutive equations.%
\begin{equation}
\partial _{t}T_{ij}^{\text{TT}}=\frac{\Lambda }{\kappa }\mathcal{E}_{ij}^{%
\text{TT}}\qquad \text{and}\qquad \varepsilon _{ikl}\partial _{k}T_{jl}^{%
\text{TT}}=-\frac{\Lambda }{\kappa }\mathcal{B}_{ij}^{\text{TT}}
\label{gravito-London equations}
\end{equation}%
These equations are directly analogous to the constitutive equations in $%
\left( \ref{London equations}\right) $ for the electric and magnetic fields,
respectively. For a sinusoidal stress tensor, we have $\partial _{t}T_{ij}^{%
\text{TT}}\propto \omega T_{ij}^{\text{TT}}$. Therefore, in the DC limit $%
\left( \omega \rightarrow 0\right) $, the first equation in $\left( \ref%
{gravito-London equations}\right) $ requires that $\mathcal{E}_{ij}^{\text{TT%
}}=0$. This implies that in the DC limit, the electric-like tensor field
vanishes completely and only the magnetic-like tensor field remains. (This
is directly analogous to the electric field vanishing throughout the entire
superconductor and only the magnetic field remaining.) Furthermore, using
the tensor fields defined in $\left( \ref{electric-like and magnetic-like
tensor fields '}\right) $ makes the wave equation in $\left( \ref{GR wave}%
\right) $ become\footnote{%
Note that the vector identity, $\nabla ^{2}\vec{A}=\nabla \left( \nabla
\cdot \vec{A}\right) -\nabla \times \left( \nabla \times \vec{A}\right) $,
is effectively used on $h_{ij}^{\text{TT}}$ in a form given by $\nabla
^{2}h_{ij}^{\text{TT}}=\partial _{i}\partial _{k}h_{kj}^{\text{TT}%
}-\varepsilon _{ikl}\varepsilon _{lmn}\partial _{k}\partial _{m}h_{jn}^{%
\text{TT}}$. Since $\partial _{i}h_{ij}^{\text{TT}}=0$, then we are left
with $\nabla ^{2}   h_{ij}^{\text{TT}}=-\varepsilon _{ikl}\partial _{k}\mathcal{%
B}_{lj}^{\text{TT}}$.}%
\begin{equation}
\varepsilon _{ikl}\partial _{k}\mathcal{B}_{lj}^{\text{TT}}=2\kappa T_{ij}^{%
\text{TT}}+\varepsilon _{\text{G}}\mu _{\text{G}}\partial _{t}\mathcal{E}%
_{ij}^{\text{TT}}  \label{gravito-Ampere for GR waves '}
\end{equation}%
This is a tensor gravito-Ampere law in the sense that a curl of a
magnetic-like tensor field is proportional to a source term plus a
time-derivative of the electric-like tensor field. It was already stated
that for sinusoidal fields and stresses, the DC limit requires $\mathcal{E}%
_{ij}^{\text{TT}}=0$. Taking a curl of $\left( \ref{gravito-Ampere for GR
waves '}\right) $ and using the second equation in $\left( \ref%
{gravito-London equations}\right) $ leads to%
\begin{equation}
\nabla ^{2}\mathcal{B}_{ij}^{\text{TT}}-2\Lambda \mathcal{B}_{ij}^{\text{TT}%
}=0
\end{equation}%
This is a Yukawa-like equation similar to $\left( \ref{Yukawa}\right) $
which implies an exponential decay solution for $\mathcal{B}_{ij}^{\text{TT}}
$ and therefore an associated penetration depth. For boundary conditions
given by $\mathcal{B}_{ij}^{\text{TT}}\left( 0\right) =\mathcal{B}_{ij,0}^{%
\text{TT}}$ and $\underset{\vec{x}\rightarrow \infty }{\lim }\mathcal{B}%
_{ij}^{\text{TT}}\left( x\right) =0$, the solution is $\mathcal{B}%
_{ij}^{\text{TT}}\left( x\right) =\mathcal{B}_{ij,0}^{\text{TT}}\left(
x\right) e^{-x/\lambda _{G}}$, where $\lambda _{\text{G}}=1/\sqrt{2\Lambda }$
with a value given by $\left( \ref{lambda_G (waves)}\right) $. This implies
that the magnetic-like tensor field is ``expelled'' in a gravitational Meissner-like effect. In the
standard Meissner effect, the expulsion of a magnetic field from a
superconductor is strong enough to cause a force on magnetic material that
allows for the levitation of magnets above superconductors. In the case of the
gravitational Meissner-like effect, the ``expulsion'' of the gravitational field could also be
interpreted as an effective ``force''
throughout the dark energy density of the universe. Since this
``force'' will exist in all directions
throughout the universe, then the manifestation of this ``force'' could be the observed outward accelerated expansion
of the universe.\vspace*{0.3in}

\section{Gravitational \textquotedblleft gauge\textquotedblright\ symmetry breaking}
The graviton becoming massive due to dark energy can also be understood as a symmetry breaking of the gravitational \textquotedblleft
gauge\textquotedblright\ (diffeomorphism)\ symmetry. Recall that a linear
coordinate transformation, $x^{\mu }\rightarrow x^{\mu }-\xi ^{\mu }$ leads
a transformation of the metric perturbation (to linear order) given by $%
h_{\mu \nu }\rightarrow h_{\mu \nu }+\partial _{\mu }\xi _{\nu }+\partial
_{\nu }\xi _{\mu }$, where $\xi ^{\mu }\ $is an arbitrary four-displacement
vector, and $\partial _{\mu }\xi _{\nu }$ is on the order of $h_{\mu \nu }$.$%
\ $However, due to dark energy, there is essentially a preferred
gravitational \textquotedblleft gauge\textquotedblright\ just as there is in
electromagnetism for a superconductor. To see this, consider the
conservation of energy-momentum-stress in linearized GR\ given by $\partial
_{\mu }T^{\mu \nu }=0$. This leads to%
\begin{equation}
\tfrac{1}{c}\dot{T}^{00}+\partial _{i}T^{0i}=0\qquad \text{and}\qquad \tfrac{%
1}{c}\dot{T}^{i0}+\partial _{j}T^{ij}=0
\end{equation}%
Taking the time-derivative of the first equation, and the divergence of
the second equation, and then combining the results leads to $\partial
_{i}\partial _{j}T^{ij}=\tfrac{1}{c^{2}}\ddot{T}^{00}$. To lowest order, $%
T^{00}=\rho _{\text{m}}c^{2}$ is the energy density. Since it is known that
the energy density of dark energy remains constant in time, then $\dot{T}%
^{00}=0$ and therefore, $\partial _{i}\partial _{j}T^{ij}=0$. Lastly, using
the constitutive equation $\left( \ref{DE constitutive equation}\right) $
and $h_{ij}\approx h^{ij}$ to linear order, gives%
\begin{equation}
\partial _{i}\partial _{j}h_{ij}^{\text{TT}}=0
\label{graviton gauge condition}
\end{equation}%
which is consistent with the fact that $h_{ij}^{\text{TT}}$ is transverse.
Notice that $\left( \ref{graviton gauge condition}\right) $ is analogous to
the London gauge, $\partial _{i}A^{i}=0$, which breaks the electromagnetic gauge symmetry in a
superconductor. A similar notion of spontaneous symmetry breaking for the gravitational field (leading to an associated Meissner-like effect in a superconductor) is also described in \cite{Agop(1),Agop(2)}. However, the effect involves the gravito-magnetic field, defined in terms of $h_{0i}$, not the gravitational wave field described in terms of $h_{ij}^{\text{TT}}$.

\bigskip
\section{A chemical potential and critical temperature for the FLRW Universe}

The full action which contains the Ginzburg-Landau (G-L) theory in curved
space-time, electromagnetism (EM)\ in curved space-time, and the
gravitational field itself, is shown in \cite{Bertschinger(GL),WeinsteinandBertschinger} as $S=\int \mathcal{L} \sqrt{-g}dx^{4}$, where
the Lagrangian density is\footnote{%
Note that $\hslash =c=1$ in \cite{WeinsteinandBertschinger}, however,
these constants will be left explicit in this formulation. Also, $\Lambda $
is not present.
\par
{}}%
\begin{eqnarray}
\mathcal{L} &=&\left( D^{\mu }\varphi \right) ^{\ast }\left( D_{\mu }\varphi
\right) +\alpha \left\vert \varphi \right\vert ^{2}+\dfrac{\beta }{2}%
\left\vert \varphi \right\vert ^{4}  \notag \\
&&  \notag \\
&&-\dfrac{1}{4\mu _{0}}F^{\mu \nu }F_{\mu \nu }+\dfrac{1}{2\kappa }\left(
R-2\Lambda \right)  \label{G-L Lagrangian density w/ GR}
\end{eqnarray}%
Note that $g\equiv \det \left( g_{\mu \nu }\right) $ is the Jacobian, and $%
\sqrt{-g}~d^{4}x$ is the invariant four-volume element. Therefore, the
action is invariant under general coordinate transformations. The terms of
the Lagrangian density can be described as follows.\bigskip

\begin{itemize}
\item The first term is a \textquotedblleft kinetic\
term,\textquotedblright\ where $\left( D^{\mu }\varphi \right) ^{\ast
}\left( D_{\mu }\varphi \right) =g^{\mu \nu }\left( D_{\mu }\varphi \right)
^{\ast }\left( D_{\nu }\varphi \right) $ involves the metric in curved
space-time, and $D_{\mu }\equiv \nabla _{\mu }-iqA_{\mu }$ is the gauge
covariant derivative. Here $A^{\mu }$ is the four-potential, and $\varphi $
is a complex scalar field, $\varphi =\varphi _{1}+i\varphi _{2}$. Since $%
\varphi $ is a scalar, then $\nabla _{\mu }\varphi =\partial _{\mu }\varphi $
and there is no need for the Christoffel symbol (Levi-Civita
connection).\bigskip

\item The second term is a \textquotedblleft mass term\textquotedblright\
since it is related to the Klein-Gordon action by setting $\alpha =k_{\text{c%
}}^{2}$, where $k_{\text{c}}\equiv mc/\hslash $ is the reduced Compton wave
number, and $mc^{2}$ is the rest mass energy.\bigskip

\item The third term (which is quartic in $\varphi $)\ represents
self-interactions.\bigskip

\item The fourth term is the electromagnetic Lagrangian density which leads
to Maxwell's equations. In curved space-time, $F^{\mu \nu }F_{\mu \nu
}=g_{\mu \kappa }g_{\nu \lambda }F^{\mu \nu }F^{\kappa \lambda }$, where $%
F_{\mu \nu }=\partial _{\mu }A_{\nu }-\partial _{\nu }A_{\mu }$ is the
electromagnetic\ field strength tensor.\footnote{%
Note that the EM\ field strength tensor in curved space-time is $F_{\mu \nu
}=\nabla _{\mu }A_{\nu }-\nabla _{\nu }A_{\mu }$. However, the connection
coefficients cancel so the covariant derivatives can be reduced to partial
derivatives.
\par
{}}\bigskip

\item The fifth term is the Einstein-Hilbert action leading to Einstein's
field equation of General Relativity.\ Here $R=g^{\mu \nu }R_{\mu \nu }$ is
the Ricci scalar, $R_{\mu \nu }=g^{\sigma \rho }R_{\mu \sigma \nu \rho }$ is
the Ricci tensor, and $R_{\hspace{0.05in}\rho \gamma \sigma }^{\mu }$ is the
Riemann tensor. Also, $\kappa \equiv 8\pi G/c^{4}$, and $\Lambda $ is the
Cosmological Constant. If there is any other matter source (besides the
charged massive scalar field), then the last term of the Lagrangian density
would be $\dfrac{1}{2\kappa }\left( R-2\Lambda \right) + \mathcal{L} _{\text{M%
}}$, where $\mathcal{L}_{\text{M}}$ is the Lagrangian density of any
classical matter field.\footnote{%
The associated stress tensor of the matter can be obtained from $T_{\mu \nu }=-2%
\frac{\delta \mathcal{L}_{\text{M}}}{\delta g^{\mu \nu }}+g_{\mu \nu
}\mathcal{L} _{\text{M}}$ or $T^{\mu \nu }=2\frac{\delta S_{\text{M}}}{%
\delta g_{\mu \nu }}\ $,\ where $S_{\text{M}}=\int \mathcal{L} _{\text{M}}%
\sqrt{-g}dx^{4}$.}\bigskip
\end{itemize}

The four-potential, scalar field, and covariant derivative each transform,
respectively, as%
\begin{eqnarray}
A_{\mu }^{\prime } &=&A_{\mu }-\nabla _{\mu }\chi  \\
&&  \notag \\
\varphi ^{\prime } &=&e^{-\frac{i}{\hslash }q\chi }\varphi  \\
&&  \notag \\
D_{\mu }^{\prime }\varphi  &=&e^{-\frac{i}{\hslash }q\chi }D_{\mu }\varphi 
\end{eqnarray}%
where $\chi $ is any real gauge function. Therefore a gauge transformation
will introduce to the scalar field a phase factor $e^{-i\phi }$, where $\phi
=\frac{q}{\hslash }\chi $ is the phase. However, $\left( D_{\mu }\varphi
\right) ^{\ast }\left( D_{\nu }\varphi \right) $ and the full action remain
invariant. Recall that the non-relativistic Schr\"{o}dinger wave function is
related to a relativistic scalar field by $\varphi =\Psi e^{-\frac{i}{%
\hslash }mc^{2}t}$. In the context of the G-L model of superconductivity, $%
\Psi $ is interpreted as a complex order parameter describing the Cooper
pairs as effectively a condensate, where $\left\vert \Psi \right\vert
^{2}=n_{\text{s}}$ is the number density of Cooper pairs. As discussed in 6.3.1 of 
\cite{Nazarov}, the transformation of the condensate wave function, $\Psi
\rightarrow e^{-\frac{i}{\hslash }q\chi }\Psi $, implies that a macroscopic
variable $\Psi $ acquires a phase upon a gauge transformation and thus
manifests symmetry breaking.\bigskip 

Notice that the standard G-L model of superconductivity is obtained by using
a time-independent wave function, a magnetic field energy density, $F^{\mu
\nu }F_{\mu \nu }=-2\vec{B}^{2}$, and neglecting gravity (the Ricci scalar
and Cosmological constant). Then writing (\ref{G-L Lagrangian density w/ GR}) as a free energy density gives\footnote{ Note that the Helmholtz free energy is found using, $F=-k_{B}T\ln (Z) ,$ where the partition function for a canonical ensemble is given by $Z=\sum\limits_{n}\exp ( -\beta E_{n}) ,$ with $E_{n}$ being the energy modes of the system, and $\beta =(k_{B} T)^{-1}$. For a superconductor, all the Cooper pairs are considered as condensed into a Bose-Einstein condensate where the particles are in the zero-momentum ground state and therefore behave as effectively a single coherent particle. In that case, the Helmholtz free energy is equivalent to the energy.}
\begin{equation}
\mathcal{F}_{\text{GL}}=\dfrac{\hslash ^{2}}{2m}\left\vert \left( \nabla -iq\vec{A}%
\right) \Psi \right\vert ^{2}+\alpha \left\vert \Psi \right\vert ^{2}+\dfrac{%
\beta }{2}\left\vert \Psi \right\vert ^{4}+\dfrac{1}{2\mu _{0}}\vec{B}^{2}
\label{GL Lagrangian density}
\end{equation}%
where $\mathcal{F}=\mathcal{F}_{{\text{n}}}+\mathcal{F}_{\text{GL}}$ is the total free enegy
which includes the normal state \cite{Tinkham}. The potential energy
contribution, $\alpha \left\vert \Psi \right\vert ^{2}+\tfrac{\beta }{2}%
\left\vert \Psi \right\vert ^{4}$, is minimized for%
\begin{equation}
\alpha +\beta \left\vert \Psi \right\vert ^{2}=0  \label{Minimized}
\end{equation}%
The expression in $\left( \ref{Minimized}\right) $ admits two solutions
which are $\left\vert \Psi \right\vert ^{2}=0$ and $\left\vert \Psi
\right\vert ^{2}=-\alpha /\beta $. Over a small range of temperatures near $%
T_{c}$, the parameters $\alpha $ and $\beta $ have the approximate values $%
\alpha \left( T\right) \approx \alpha _{0}\left( \frac{T}{T_{c}}-1\right) $,
and $\beta \left( T\right) \approx \beta _{0}$, where $T_{c}$ is the
critical temperature, and $\alpha _{0}$, $\beta _{0}$ are both defined as
positive constants \cite{Poole}. If $T<T_{c}$, then $\alpha <0$ and $%
\left\vert \Psi \right\vert ^{2}=-\alpha /\beta $ is positive which
corresponds to the fact that in G-L theory, $\left\vert \Psi \right\vert
^{2}=n_{s}$ is the number density of Cooper pairs which can only be
positive. If $T>T_{c}$, then $\alpha >0$. However, since $\left\vert \Psi
\right\vert ^{2}$ cannot be negative, then we must have $\left\vert \Psi
\right\vert ^{2}=0$, which corresponds to the destruction of the
superconducting state. Hence, when the temperature crosses between $T>T_{c}$
and $T<T_{c}$, then $\left\vert \Psi \right\vert $ spontaneously switches
between zero and $\left\vert \Psi \right\vert ^{2}=-\alpha /\beta $. This is
the well-known process of spontaneous symmetry breaking for a
superconductor.\bigskip

Furthermore, the chemical potential of the superconducting phase is defined
as the variation of the free energy with respect to the number of Cooper
pairs of the condensate: $\mu \equiv \delta \mathcal{F}/\delta n_{\text{s}}$%
. The energy of a Cooper pair is $\mu =2\mu _{\text{e}}$, twice the electron
chemical potential. Since the time dependence of an energy eigenstate in
quantum mechanics comes from the phase factor, $\exp \left( -iEt/\hslash
\right) $, which for the order parameter, $\Psi =\left\vert \Psi \right\vert
e^{-i\phi }$, is $\exp \left( -i\mu t/\hslash \right) $, then the chemical
potential is%
\begin{equation}
\mu =\dfrac{\delta \mathcal{F}}{\delta n_{\text{s}}}=-\hslash \frac{d\phi }{dt}
\end{equation}%
Using $\left( \ref{GL Lagrangian density}\right) $ leads to%
\begin{equation}
\mu =\alpha +\beta \left\vert \Psi \right\vert ^{2}
\label{chemical potential}
\end{equation}%
Therefore, $\left( \ref{Minimized}\right) $ corresponds to $\mu =0$ as
the minimum chemical potential of the system.\bigskip

We can return to $\left( \ref{G-L Lagrangian density w/ GR}\right) $ and
exclude electromagnetism $\left( A^{\mu }=0\right) $ to obtain a G-L
free energy density for the case of dark energy. Note that in curved
space-time, the coordinate volume is expressed in terms of the proper volume
as $dV=dV_{\text{proper}}/\sqrt{-g}$. Then the G-L free energy density with respect to
proper volume is\footnote{%
Note that describing the effects of dark energy with $\left( \ref{GL free
energy with gravity}\right) $ is similar to the concept of Quintessence
which is an attempt to describe dark energy with the help of a scalar
field $\varphi $ which can indeed lead to negative pressure. See \cite{Xenos}%
-\cite{Ratra}.}%
\begin{eqnarray}
\mathcal{F}_{\text{GL}} &=&\sqrt{-g}\left[ \dfrac{\hslash ^{2}}{2m}%
\left\vert \nabla \Psi \right\vert ^{2}+\alpha \left\vert \Psi \right\vert
^{2}+\dfrac{\beta }{2}\left\vert \Psi \right\vert ^{4}\right.   \notag \\
&&  \notag \\
&&\left. +\dfrac{1}{2\kappa }\left( R-2\Lambda \right) \right] 
\label{GL free energy with gravity}
\end{eqnarray}

To describe the case of dark energy, the metric of the universe can be
written as the Robertson-Walker metric \cite{Friedman} with zero curvature $%
\left( k=0\right) $ which is%
\begin{equation}
g_{00}=-1,\qquad g_{0i}=0,\qquad g_{ij}=a^{2}\delta _{ij}
\label{FLRW metric components}
\end{equation}%
The metric is written in terms of cosmological time coordinate $t$, and
spatial coordinates $x$, $y$, $z$, where $a(t)$ is the cosmological scale
factor. For the metric in $\left( \ref{FLRW metric components}\right) $, the Ricci
scalar and Jacobian are respectively,%
\begin{equation}
R=\dfrac{6}{c^{2}}\left( \dfrac{\ddot{a}}{a}+\dfrac{\dot{a}^{2}}{a^{2}}%
\right) \qquad \text{and}\qquad g=-a^{6}  \label{Ricci scalar and Jacobian}
\end{equation}%
Therefore $\left( \ref{GL free energy with gravity}\right) $ becomes%
\begin{eqnarray}
\mathcal{F}_{\text{GL}}^{\left( \text{FLRW}\right) } &=&\dfrac{\hslash
^{2}a^{3}}{2m}\left\vert \nabla \Psi \right\vert ^{2}+a^{3}\left( \alpha
\left\vert \Psi \right\vert ^{2}+\dfrac{\beta }{2}\left\vert \Psi
\right\vert ^{4}\right)   \notag \\
&&  \notag \\
&&+\dfrac{3a^{3}}{\kappa c^{2}}\left( \dfrac{\ddot{a}}{a}+\dfrac{\dot{a}^{2}%
}{a^{2}}\right) -\dfrac{a^{3}\Lambda }{\kappa }
\label{GL free energy with gravity for FLRW}
\end{eqnarray}%
The chemical potential that includes the role of dark energy now becomes
\begin{equation}
\mu ^{\left( \text{FLRW}\right) }=a^{3}\left( \alpha +\beta \left\vert \Psi
\right\vert ^{2}\right)   \label{mu_DE^(FLRW)}
\end{equation}%
which is clearly related to $\left( \ref{chemical potential}\right) $ by $%
\mu ^{\left( \text{FLRW}\right) }=a^{3}\mu $. However, notice
the potential energy is still minimized according to $\left( \ref{Minimized}%
\right) $ as before. Therefore, spontaneous symmetry breaking still occurs
when the temperature crosses from $T>T_{c}$ to $T<T_{c}$, and $\left\vert
\Psi \right\vert $ switches from zero to $\left\vert \Psi
\right\vert ^{2}=-\alpha /\beta $.\bigskip 

Now consider the role of a plane-fronted gravitational wave propagating
through the background of an FLRW\ universe. Then the metric in $\left( \ref%
{FLRW metric components}\right) $ becomes%
\begin{equation}
g_{00}=-1,\qquad g_{0i}=0,\qquad g_{ij}=a^{2}\delta _{ij}+h_{ij}^{TT}
\end{equation}%
In that case, the Ricci scalar is still the same, but the determinant of the
metric becomes $g=-a^{6}\left( 1-h_{\oplus }^{2}\right) +a^{2}h_{\otimes
}^{2}$, where $h_{\oplus }$ is for plus-polarization, and $h_{\otimes }$ is
for cross-polarization. To first order in the wave amplitude, we can
approximate%
\begin{equation}
\sqrt{-g}=a^{3}\sqrt{1-\left( h_{\oplus }^{2}+h_{\otimes }^{2}/a^{4}\right) }%
\approx a^{3} -\tfrac{1}{2a}\left( a^{4}h_{\oplus }^{2}+h_{\otimes
}^{2}\right) 
\end{equation}%
Then the G-L free energy density in $\left( \ref{GL free energy with gravity
for FLRW}\right) $ has an additional term added to it which is associated
with the gravitational wave (GW). This leads to%
\begin{eqnarray}
\mathcal{F}_{\text{GL}}^{\left( \text{FLRW+GW}\right) } &=&\mathcal{F}_{%
\text{GL}}^{\left( \text{FLRW}\right) }-\dfrac{1}{2a}\left( a^{4}h_{\oplus
}^{2}+h_{\otimes }^{2}\right) \left[ \dfrac{\hslash ^{2}}{2m}\left\vert
\nabla \Psi \right\vert ^{2}\right.   \notag \\
&&  \notag \\
&&\left. +\left( \alpha \left\vert \Psi \right\vert ^{2}+\dfrac{\beta }{2}%
\left\vert \Psi \right\vert ^{4}\right) \right] 
\end{eqnarray}%
Then $\left( \ref{mu_DE^(FLRW)}\right) $ becomes%
\begin{equation}
\mu ^{\left( \text{FLRW+GW}\right) }=a^{3}\left[ 1-\tfrac{1}{2a}\left(
a^{4}h_{\oplus }^{2}+h_{\otimes }^{2}\right) \right] \left( \alpha +\beta
\left\vert \Psi \right\vert ^{2}\right) 
\end{equation}%
In this case, the potential energy is minimized when
\begin{equation}
\left[ 1-\tfrac{1}{2a}\left( a^{4}h_{\oplus }^{2}+h_{\otimes }^{2}\right) %
\right] \left( \alpha +\beta \left\vert \Psi \right\vert ^{2}\right) =0
\end{equation}%
Although the gravitational wave has altered the chemical potential,
nevertheless the condition necessary for spontaneous symmetry-breaking
still follows $\left( \ref{Minimized}\right) $ as before.\bigskip 

\section{An Unruh--Hawking-like effect from Riemann Normal Coordinates}

An alternative approach is described in \cite{Atanasov} which uses Riemann
Normal Coordinates (RNC). As shown in \cite{Poisson}, RNC (to second order
in the coordinates)\ can be written as%
\begin{equation}
g_{\mu \nu }=\eta _{\mu \nu }-\tfrac{1}{3}R_{\mu \sigma \nu \rho }x^{\sigma
}x^{\rho }  \label{RNC}
\end{equation}%
Then the inverse metric is
\begin{equation}
g^{\mu \nu }=\eta ^{\mu \nu }-\tfrac{1}{3}g^{\mu \lambda }g^{\nu \gamma
}R_{\lambda \sigma \gamma \rho }x^{\sigma }x^{\rho }
\end{equation}%
and the kinetic term in $\left( \ref{G-L Lagrangian density w/ GR}\right) $
becomes%
\begin{equation}
\left( D^{\mu }\varphi \right) ^{\ast }\left( D_{\mu }\varphi \right)
=\left( \eta ^{\mu \nu }-\tfrac{1}{3}g^{\mu \lambda }g^{\nu \gamma
}R_{\lambda \sigma \gamma \rho }x^{\sigma }x^{\rho }\right) \partial _{\mu
}\partial _{\nu }\varphi   \label{kinetic term in RNC}
\end{equation}%
It is immediately evident that the kinetic energy now includes an additive
term that involves the curvature via the Riemann tensor, $R_{\lambda \sigma
\gamma \rho }$. As will be shown below, this added term can be interpreted
as a \textquotedblleft gravitational potential energy\textquotedblright\
which will modify the chemical potential.\bigskip 

It is stated in \cite{Atanasov} that when the 4-D space-time can be split
into 3+1 dimensions, the induced Riemannian metric $g_{ij}$ on the 3-D
hyper-plane can be used to write the Laplace-Beltrami operator $\nabla _{%
\text{LB}}^{2}$ acting on a scalar as $\nabla _{\text{LB}}^{2}\Psi =\frac{1}{%
\sqrt{g}}\partial _{i}\left( \sqrt{g}g^{ij}\partial _{j}\Psi \right) $.
Since $\left\vert \Psi \right\vert ^{2}=n_{\text{s}}$ is a particle density, then
the particle density with respect to proper volume is $\left\vert \Psi
\right\vert ^{2}/\sqrt{-g}$. To lowest order in $x^{i}$, this leads to%
\begin{equation}
\nabla _{\text{LB}}^{2}\frac{\Psi }{\left( -g\right) ^{1/4}}=\nabla ^{2}\Psi
+\frac{1}{12}R^{\left( \text{3D}\right) }\Psi 
\end{equation}%
where $R^{\left( \text{3D}\right) }\equiv \frac{3}{4}R$ is the induced 3-D
Ricci scalar curvature. Then $\left( \ref{GL free energy with gravity}%
\right) $ using RNC becomes%
\begin{eqnarray}
\mathcal{F}_{\text{GL}}^{\left( \text{RNC}\right) } &=&\dfrac{\hslash ^{2}}{%
2m}\left\vert \nabla \Psi \right\vert ^{2}+\frac{\hslash ^{2}}{24m}R^{\left( 
\text{3D}\right) }\left\vert \Psi \right\vert ^{2}  \notag \\
&&+\sqrt{-g}\left[ \alpha \left\vert \Psi \right\vert ^{2}+\dfrac{\beta }{2}%
\left\vert \Psi \right\vert ^{4}\right.   \notag \\
&&+\left. \dfrac{1}{2\kappa }\left( R-2\Lambda \right) \right] 
\label{GL free energy with gravity in RNC}
\end{eqnarray}%
As previously mentioned, the second term is an effective gravitational
potential energy which was essentially pulled out of the kinetic energy via
use of RNC. Again using $\left( \ref{Ricci scalar and Jacobian}\right) $
leads to%
\begin{eqnarray}
\mathcal{F}_{\text{GL}}^{\left( \text{FLRW, RNC}\right) } &=&\dfrac{\hslash
^{2}}{2m}\left\vert \nabla \Psi \right\vert ^{2}+\dfrac{3\hslash ^{2}}{%
16mc^{2}}\left( \dfrac{\ddot{a}}{a}+\dfrac{\dot{a}^{2}}{a^{2}}\right)
\left\vert \Psi \right\vert ^{2}  \notag \\
&&  \notag \\
&&+a^{3}\left( \alpha \left\vert \Psi \right\vert ^{2}+\dfrac{\beta }{2}%
\left\vert \Psi \right\vert ^{4}\right) +\cdots 
\label{GL free energy (FLRW, RNC)}
\end{eqnarray}%
The chemical potential in this approach now becomes
\begin{equation}
\mu ^{\left( \text{FLRW, RNC}\right) }=\dfrac{3\hslash ^{2}}{16mc^{2}}\left( 
\dfrac{\ddot{a}}{a}+\dfrac{\dot{a}^{2}}{a^{2}}\right) +a^{3}\left( \alpha
+\beta \left\vert \Psi \right\vert ^{2}\right)   \label{mu_(FLRW, RNC)}
\end{equation}%
The condition for spontaneous symmetry breaking is now obtained by setting $%
\mu ^{\left( \text{FLRW, RNC}\right) }=0$. Also using $\alpha \left(
T\right) \approx \alpha _{0}\left( \frac{T}{T_{\text{c}}}-1\right) $ leads to%
\begin{equation}
\left\vert \Psi \right\vert ^{2}=-\frac{\alpha _{0}}{\beta }\left( \frac{T}{%
T_{\text{c}}}-1\right) -\dfrac{3\hslash ^{2}\left( a\ddot{a}+\dot{a}%
^{2}\right) }{16mc^{2}a^{5}\beta }
\end{equation}%
Notice that $T<T_{c}$ is no longer sufficient to make $\left\vert \Psi
\right\vert ^{2}>0$ for symmetry breaking. Instead, we must have%
\begin{equation}
T<\left( 1-\gamma \right) T_{\text{c}}\qquad \text{where}\qquad \gamma
\equiv \dfrac{3\hslash ^{2}\left( a\ddot{a}+\dot{a}^{2}\right) }{%
16mc^{2}a^{5}\alpha _{0}}  \label{T < gamma*T_c}
\end{equation}%
This means that the rate of expansion of the universe ($\ddot{a}$ and $\dot{a%
}^{2}$) determines $\gamma $ and hence determines the effective critical
temperature, $\gamma T_{c}$, for spontaneous symmetry breaking.\bigskip 

Comparing $\left( \ref{mu_DE^(FLRW)}\right) $ and $\left( \ref{mu_(FLRW,
RNC)}\right) $, it is evident that using RNC\ introduces an interesting
modification to the chemical potential. Also, $\left( \ref{T < gamma*T_c}%
\right) $ introduces a modification to the condition for spontaneous
symmetry breaking. These modifications can be understood by considering the
mathematical and physical meaning of using RNC.

Mathematically, RNC\ use geodesics through a given point to define the
coordinates for nearby points. If $t^{\mu }$ is the unit tangent vector to a
geodesic at a given point $O$, and $s$ is the geodesic arc length measured
from $O$ to a point $P$, then the RNC of $P$ are defined to be $x^{\mu
}=st^{\mu }$. Since $t^{\mu }$ is constant along each geodesic through $O$,
then the Christoffel symbol (Levi-Civita connection) is $\Gamma _{\mu \nu
}^{\sigma }=0$ at $O$. Then using the geodesic equation of motion,%
\begin{equation}
\dfrac{d^{2}x^{\mu }}{ds^{2}}+\Gamma _{\rho \sigma }^{\mu }\dfrac{dx^{\rho }%
}{ds}\dfrac{dx^{\sigma }}{ds}=0
\end{equation}%
leads immediately to $\dfrac{d^{2}x^{\mu }}{ds^{2}}=0$ at $O$. This means
that in RNC, the coordinate system is chosen such that it forms a local
inertial frame at a specific point in spacetime \cite{Brewin}. This is
evident by the first term in $\left( \ref{RNC}\right) $ which is $\eta _{\mu
\nu }$. Furthermore, the second term (involving $R_{\lambda \sigma \gamma
\rho }$) implies that curvature is sampled when moving from $O$ to $P$.
However, by the Equivalence Principle, a \textit{single} observer cannot
differentiate between a local acceleration and a sampling of spacetime
curvature.

Physically, this means that if an observer accelerates along their
worldline, this acceleration could be perceived as a contribution to the
curvature of spacetime in RNC. This is consistent with the fact that an
additional contribution involving the Riemann tensor appears in the kinetic
energy $\left( \ref{kinetic term in RNC}\right) $. It is also consistent
with the \textit{induced} Riemannian metric $g_{ij}$ on the 3-D hyper-plane,
and the \textit{induced} 3-D Ricci scalar curvature, $R^{\left( \text{3D}%
\right) }$, in the free energy density $\left( \ref{GL free energy with
gravity in RNC}\right) $. However, the additional term in these expressions
may not necessarily be due to actual space-time curvature but rather
non-inertial acceleration on a wordline.

For the case of the FLRW\ metric, the use of RNC leads to the terms
involving $\ddot{a}$ and $\dot{a}^{2}$ in $\left( \ref{GL free energy (FLRW,
RNC)}\right) -\left( \ref{T < gamma*T_c}\right) $. Hence, for an observer
accelerating on their wordline, there is a modified effective chemical
potential, and a modified critical temperature for symmetry breaking to
occur. Interestingly, comparing $\left( \ref{mu_DE^(FLRW)}\right) $ and $%
\left( \ref{mu_(FLRW, RNC)}\right) $ shows that the chemical potential
becomes \textit{higher} using RNC. However, from $\left( \ref{T < gamma*T_c}%
\right) $ it is evident that because $\gamma <1$, then the effective
critical temperature, $\gamma T_{\text{c}}$, is \textit{lower} than $T_{%
\text{c}}$. Therefore, it appears that $T$ must fall lower than before in
order to achieve symmetry breaking since it requires $T<\gamma T_{\text{c}}$%
.

However, rearranging $\left( \ref{T < gamma*T_c}\right) $ as $T+\gamma T_{%
\text{c}}<T_{\text{c}}$ leads to an alternative interpretation. The left
side of the inequality indicates that an observer accelerating on their
worldline would observe an effective temperature given by $T_{\text{eff}%
}=T+\gamma T_{\text{c}}$ which is \textit{higher} than an inertial observer
who observes a temperature $T$. Therefore, symmetry breaking for the
accelerated observer requires $T_{\text{eff}}<T_{\text{c}}$, while symmetry
breaking for the inertial observer requires $T<T_{\text{c}}$. From this
point of view, the condition for symmetry breaking is the \textit{same} for
both observers (namely, it requires reaching a temperature below $T_{\text{c}%
}$), but the accelerated observer must drop in temperature more than the
inertial observer because the accelerated observer has an additional
temperature gap of $\gamma T_{\text{c}}$. This interpretation is consistent
with the fact that $\gamma $ can be traced back to the additional
contribution to the kinetic energy involving the Riemann tensor in $\left( %
\ref{kinetic term in RNC}\right) $. The higher kinetic energy is ultimately
the reason for the higher observed temperature, $T_{\text{eff}}$.

The idea of observing a higher temperature for an accelerated observer can
be likened to the Unruh effect. Recall that the Unruh effect demonstrates
that an observer who is uniformly accelerating through empty space will
perceive a thermal bath and associated temperature. In contrast, an inertial
observer in the same region of spacetime would observe no temperature.
Similarly, using RNC\ leads to an Unruh-like effect where the observer in an
FLRW\ universe observes a temperature shift given by $T_{\text{eff}%
}=T+\gamma T_{c}$. In fact, the Unruh temperature is known to be $T_{\text{U}%
}=\dfrac{\hslash a}{2\pi ck_{\text{B}}}$, where $a$ is a proper acceleration.
Similarly, the Hawking temperature is also known to be $T_{\text{H}}=\dfrac{%
\hslash g}{2\pi ck_{\text{B}}}$, where $g$ is the surface gravity of a black
hole. In light of the Equivalence Principle, the two temperatures are
sometimes referred to as the Hawking--Unruh temperature \cite{Alsing}.
Similarly, in the context of this paper, the temperature shift of $T_{\text{%
eff}}=T+\gamma T_{\text{c}}$ could be either due to acceleration (like the case for the
Unruh effect)\ or space-time curvature (like the case for Hawking radiation).

Another approach that can be used instead of RNC is Fermi normal coordinates%
\footnote{%
It is argued in \cite{Parker(1)} that Fermi normal coordinates are
appropriate for a problem involving energy levels, in contrast to Riemann
normal coordinates.} which also expresses the Hamiltonian in terms of the
Riemann curvature tensor \cite{RB,BR,Parker(1),Parker(2),Parker(3),Minter}. In fact, there exist many other approaches for describing the
coupling of gravity to the Ginzburg-Landau scalar field model \cite{Lano(1),CP,PT,GP,CCJ,BD}. A variety of other possibilities is
also discussed in \cite{SCMeissnereffects}.


\section{Conclusion}
Guided by the concept of photons becoming massive in a superconductor, in this paper we demonstrate that gravitons are also massive due to dark energy acting like a superconducting medium. The superconductor analogy can be justified from a quantum gravity point of view. Namely, at Planck length distances, the spacetime is believed to be discrete (``spacetime quanta'') that may take the shape of 24-cell \cite{FaragAli:2024jpo} hence the analogy with the atomic lattice of the superconductor.  It is also expected that this structure is dynamic in the sense that quantum fluctuations can generate phonon-like particles that can interact with gravitons. The key idea is that the constitutive equation describing the response of a superconductor to an electromagnetic field (the London equation) is formally analogous to a constitutive equation obtained from the Einstein field equation with a cosmological constant. This constitutive equation also has formal similarities to a constitutive equation in elastic mechanics involving a bulk modulus and shear modulus. These moduli can also be interpreted as gravitational conductivity with a value that aligns with the energy density of dark energy.

When the constitutive equation described above is used in the vacuum Einstein field equation, it naturally leads to a screening length scale for the gravitational scalar potential which has a Yukawa solution. Hence the value of the screening length is determined on various scales according to the dominance of dark energy (on cosmological scales) or dark matter (on galactic scales).

For the case of gravitational waves, it is found that the screening length scale is frequency-dependent. The DC limit leads to a minimum screening length on the order of the size of the universe and a corresponding maximum graviton mass. Several other properties are found for gravitational waves propagating through the dark energy medium such as a plasma frequency, index of refraction, and impedance. It is shown that very low-frequency gravitational waves (such as $\omega \sim 10^{-17}~$rad/s) will propagate 0.5\% slower than light. There is also a predicted Meissner-like effect where the gravitational wave tensor field is \textquotedblleft expelled\textquotedblright\ due to dark energy. This might be interpreted in terms of the expansion of the universe as dark energy causes an
outward ``expulsion'' of space-time similar to a superconductor expelling a magnetic field.

Finally, it was shown that the fundamental cause of these effects can be interpreted as a type of spontaneous symmetry breaking of a scalar field similar to the Higgs mechanism. There is an associated chemical potential and critical temperature associated with the symmetry breaking. In fact, it was shown using Riemann's normal coordinates that there is an induced chemical potential that can be interpreted as a genuine space-time curvature or a result of the acceleration of a non-inertial observer. Consequently, there is an associated shift in the critical temperature which can be described as a Hawking-Unruh effect.\\

 While our model provides an effective explanation for dark matter and dark energy say, in galactic and cosmological scales, it remains however an effective model rather than a fundamental theory. This means that the mysterious nature of dark matter and dark energy still persists and a more fundamental theory of quantum-gravity is needed to explain the nature of dark matter and dark energy. Recent works suggest that incorporating an invariant minimum speed at the quantum level, as proposed in Symmetrical Special Relativity (SSR), might offer a foundational principle for quantum gravity. Future research should explore these perspectives to better understand the true nature of dark matter and dark energy. For further context, see \cite{Cruz_2023,Cruz_2016,Cruz_2020}.



\end{document}